\documentclass[12pt]{article}

\usepackage{amsfonts}
\usepackage{amssymb}
\usepackage{amsthm}
\usepackage[fleqn]{amsmath}
\usepackage[mathcal]{euscript}
\usepackage{mathrsfs}

\newtheorem{con}{Conjecture}
\newtheorem{cor}{Corollary}
\newtheorem{lem}{Lemma}
\newtheorem{prp}{Proposition}
\newtheorem{thm}{Theorem}
\newtheorem{rem}{Remark}

\newcommand{\Ref}[1]{(\ref{#1})}

\newcommand{\vect}[1] {\boldsymbol{{ #1}} }

\newcommand{\pV}{{\vect{p}}}           
\newcommand{\qV}{{\vect{q}}}           

\DeclareMathAlphabet{\mathpzc}{OT1}{pzc}{m}{it}

\newcommand\pzcC{{\mathpzc{C}}}
\newcommand\pzcE{{\mathpzc{E}}}
\newcommand\pzcF{{\mathpzc{F}}}
\newcommand\pzcH{{\mathpzc{H}}}
\newcommand\pzcI{{\mathpzc{I}}}
\newcommand\pzcK{{\mathpzc{K}}}

\newcommand\pzcQ{{\mathpzc{Q}}}

\newcommand\pzcS{{\mathpzc{S}}}
\newcommand\pzcV{{\mathpzc{V}}}

\newcommand{\funcC}{\pzcC}
\newcommand{\funcH}{\pzcH}
\newcommand{\funcI}{\pzcI}
\newcommand{\funcK}{\pzcK}
\newcommand{\funcQ}{\pzcQ}

\newcommand{\uli}[1]{\underline #1 }

\newcommand{\Prob}{\mathrm{Prob}}
\newcommand{\Meas}{\mathrm{Meas}}

\newcommand{\eps}{\epsilon}
\newcommand{\veps}{\varepsilon}

\newcommand\frake{{\mathfrak{e}}}

\newcommand{\dKR}{d_{\mathrm{KR}}}

\newcommand{\dd}{\mathrm{d}}
\newcommand{\ddn}{\mathrm{d}^{^{\!3n}}\!\!\!\!}
\newcommand{\ddN}{\mathrm{d}^{^{3\powN}}\!\!\!\!}

\newcommand{\withNto}{\stackrel{\textrm{\tiny N}\to\infty}{\longrightarrow}}

\newcommand{\mun}{{\mu}_n}

\newcommand{\dotmu}{{\dot\mu}}

\newcommand{\muEINS}{{\mu}_1}
\newcommand{\muZWEI}{{\mu}_2}

\newcommand{\varrhok}{{\varrho}_k}

\newcommand{\varrhon}{{\varrho}_n}

\newcommand{\Nset}{\mathbb{N}}
\newcommand{\Rset}{\mathbb{R}}
\newcommand{\Sset}{\mathbb{S}}

\newcommand{\Csp}{\mathfrak{C}}
\newcommand{\Hsp}{\mathfrak{H}}
\newcommand{\Lsp}{\mathfrak{L}}
\newcommand{\Msp}{\mathfrak{M}}
\newcommand{\Psp}{\mathfrak{P}}

\newcommand{\cE}{{\cal E}}
\newcommand{\cH}{{\cal H}}
\newcommand{\cN}{{\cal N}}

\newcommand{\Bose}{\mathscr{Bose}} 
 
\newcommand{\Ferm}{\mathscr{Ferm}}

\newcommand{\BO}{{\infty}} 

\newcommand{\PowN}{{\scriptstyle{N}}}
\newcommand{\powN}{{\scriptscriptstyle{N}}}

\newcommand{\powZ}{{\scriptscriptstyle{Z}}}

\newcommand{\sss}{\scriptscriptstyle}
\newcommand{\tst}{\textstyle}

\newcommand{\tfrhalf}{{\textstyle{\frac{1}{2}}}}
\newcommand{\tfrHALF}{{\textstyle{\frac{1}{2}}}}

\newcommand{\tfrn}{{\tst{\frac{1}{n}}}}
\newcommand{\tfrN}{{\tst{\frac{1}{N}}}}

\newcommand{\sfrTHREEhalf}{{\scriptstyle{\frac{3}{2}}}}

\begin{document}

\title{The Hartree limit of Born's ensemble for \\
	the ground state of a bosonic atom or ion$^*$}

\vspace{-0.3cm}
\author{\normalsize \sc{Michael K.-H. Kiessling}\\[-0.1cm]
	\normalsize Department of Mathematics, Rutgers University\\[-0.1cm]
	\normalsize Piscataway NJ 08854, USA}
\vspace{-0.3cm}
\date{$\phantom{nix}$}
\maketitle
\vspace{-1.6cm}

\begin{abstract}
\noindent
	The non-relativistic bosonic ground state is studied for quantum $N$-body systems with  
Coulomb interactions, modeling atoms or ions made of $N$ ``bosonic point electrons'' bound to an 
atomic point nucleus of $Z$ absolute ``electron'' charges, treated in Born--Oppenheimer approximation 
(the nuclear mass $M=\infty$).
	By adapting an argument of Hogreve, it is shown that the (negative) $\Bose$osonic ground state energy 
$\cE_{\BO}^{\Bose}(Z,N)$ yields the monotone non-decreasing function $N\mapsto \cE_{\BO}^{\Bose}(\lambda N,N)/N^3$
for any $\lambda>0$.
	The main part of the paper furnishes a proof that whenever $\lambda \geq \lambda_*\approx 1/1.21$, 
then the limit $\veps(\lambda):=\lim_{N\to\infty}\cE_{\BO}^{\Bose}(\lambda N,N)/N^3$
is governed by Hartree theory, and the rescaled bosonic ground state wave function factors into an 
infinite product of identical one-body wave functions determined by the Hartree equation.
	The proof resembles the construction of the thermodynamic mean-field limit of the classical 
ensembles with thermodynamically unstable interactions, except that here the ensemble is Born's,
with $|\psi|^2$ as ensemble probability density function on $\Rset^{3N}$, with the Fisher information 
functional in the variational principle for Born's ensemble playing the role of the negative of the 
Gibbs entropy functional in the free-energy variational principle for the classical petit-canonical 
configurational ensemble. 
\end{abstract}
\bigskip
\bigskip

\vfill
\hrule
\smallskip\noindent
{\small 
Typeset in \LaTeX\ by the author. Revised version: August 26, 2012. 

\smallskip\noindent
$^*$ To Elliott H. Lieb on his 80th birthday, in admiration.

\smallskip\noindent
\copyright 2012 The author. This preprint may be reproduced for noncommercial purposes.}

\section{Introduction}
\vskip-0.45truecm
\noindent
	In Spring 2011 Brookhaven National Laboratory announced
the discovery of the first dozen and a half anti-$\alpha$ particles\footnotemark\
\cite{STAR}.
	Since $\alpha$ particles in their ground state have spin zero, the same can be expected
for anti-$\alpha$ particles which, therefore, will be twice negatively charged bosons for practically 
all quantum-mechanical low-energy phenomena in which they participate.
	In particular, it doesn't take much imagination now to predict that lab-produced atoms with such bosonic 
``electrons'' in place of the usual fermionic electrons are eventually going to become an experimental reality,
although the path to their controlled production and storage is cluttered with enormous technical obstacles that 
need to be overcome.
	Yet such a feat should be possible, as demonstrated by the recent success story about storing
anti-$H$ atoms \cite{HHetAL}, and by the recent lab creation of ``protonium'' \cite{protoniumA,protoniumB}, 
an atom made of a proton and an anti-proton, both fermions.

        The simplest atom formed with an anti-$\alpha$ particle as electron substitute, and presumably
the first one to be produced in the lab, would seem to be ``alphium,'' i.e. the $\alpha$ particle analog of 
protonium and of the familiar positronium (which I'd much rather like to see called ``electronium'').
	Alas, although an (anti-$\alpha$,$\alpha$) bound state is the simplest special case of an atom 
with bosonic ``electrons,'' a bosonic atom having just $N=1$ bosonic ``electrons'' cannot
display any of the effects which depend on the permutation symmetry of bosonic many-body wave functions.
	To see those, laboratory-produced atoms with $\alpha$-bosonic ``electrons'' would have to be formed 
with $N>1$ bosonic anti-$\alpha$ particles attracted by a conventional nucleus of charge $2eN$.
        So the atom with $N=2$ $\alpha$-bosonic ``electrons'' would be the simplest truly bosonic atom; 
it'll have a Beryllium nucleus, the stable isotope of which (${}^{9}$Be) is a fermion with spin $3/2$.
	In principle it is conceivable to go up to $N\approx 46$, requiring a Uranium nucleus, although it would seem 
exceedingly difficult to strip away all electrons of the atomic hull surrounding any heavier nucleus.

	Of course, alphium and the heavier bosonic atoms with anti-$\alpha$ particles as electron substitute
will be short-lived, their  anti-$\alpha$ particles annihilating with (parts of) their nuclei in a complicated 
fashion, thus opening an interesting new venue for studying the strong interactions at their lower energies.
	However, if positronium and protonium are any indicators, then alphium and the heavier $\alpha$-bosonic 
atoms may well have lifespans of the order of $\mu$-seconds.
	This is long enough to gather empirical data about electromagnetic transitions between their excited atomic 
states and their atomic ground state, for which the treatment of anti-$\alpha$ particles as bosonic ``electrons'' is 
vindicated.

\vskip-1truecm
\footnotetext{This announcement came very much to the delight of the author, who in 2009 \cite{KieJSPb} wrote:
	``Since fermionic anti-${}^3$He nuclei have already been produced in heavy ion collisions at CERN
	\cite{AntiHeDREIa,AntiHeDREIb,AntiHeDREIc}, it seems a safe bet to predict that also bosonic 
	anti-$\alpha$ particles are going to be produced in the laboratory,...'', but who had no idea that
	this was such a timely assessment of the experimental situation.}

\newpage

	Data on electromagnetic transitions in these $\alpha$-bosonic atoms should not only yield information
which depends on the symmetry of many-body bosonic wave functions, but also information about the structure of 
the anti-$\alpha$ particles themselves, for these are not at all point-like, in contrast to the conventional 
fermionic electrons.
	For instance, the fine structure of the $\alpha$-bosonic atomic spectrum should show some novel effects due
to electric quadrupole-quadrupole coupling, which is absent in atoms with true electrons.

	No later than when atoms with anti-$\alpha$ particles as bosonic ``electrons'' 
are being produced in the laboratory will theoretical physicists get busy calculating the various annihilation channels
(a challenging task) and the electromagnetic spectra of these bosonic atoms (a more ``straightforward'' 
task).
	The calculation of the bosonic spectra will require accurate numerical approximation methods, but very
likely also the traditional Hartree approximation combined with perturbation theory for Schr\"odinger operators.
	Happy news for mathematical physicists who have been busy with rigorous studies of such atoms
for some time already; see \cite{BenguriaLieb}, \cite{Lieb}, \cite{Solovej}, \cite{BachA}, 
\cite{BachEtAl}, \cite{Ruskai},  \cite{BaumgSeiringer}, \cite{KieJSPb}, \cite{Hog}, and in
particular \cite{LiebSeiringer} and references therein.

	In this work we will concern ourselves with the Hartree approximation to
the ground state of a non-relativistic atom or ion with bosonic ``electrons.'' 

	Informally, we can summarize our results as follows: let $\cE_{\BO}^{\Bose}(Z,N)$ 
with $Z=\lambda N$ denote the $\Bose$osonic ground state energy of a Born--Oppenheimer atom ($\lambda=1$) 
or ion ($\lambda\neq 1$); the index ``$_\infty$'' stands for the infinitely massive nucleus.
	Let $\psi^{(\powN)}_\lambda$ denote the (suitably scaled) bosonic ground state wave function 
of this $N$-body system, $\funcH_\lambda(\phi)$ the pertinent asymptotic ($N$-independent) Hartree functional 
of a one-point wave function, and $\uli\Delta^{(n)}_N$ the normalized empirical $n$-particle measure (as obtained
in an ideal measurement of the $N$ positions).
	We will show that whenever $\lambda\geq \lambda^*\approx 1/1.21$,~then: 

\hskip7pt $(i)$ $N^{-3}\cE_{\BO}^{\Bose}(\lambda N,N) \nearrow \inf_\phi \funcH_\lambda(\phi)$ when $N\to\infty$;

\hskip3.5pt $(ii)$ $\funcH_\lambda(\phi)$ has a unique minimizer, $\phi_\lambda$, modulo sign; 

	$(iii)$ $\psi^{(\powN)}_\lambda \to \phi_\lambda^{\otimes\Nset}$ as $N\to\infty$;

\hskip1pt $(iv)$ $\uli\Delta^{(n)}_N \to |\phi_\lambda|^2\otimes \cdots\otimes|\phi_\lambda|^2\ (n$-fold product), 
	as $N\to\infty$.

\noindent
	The precise statements are given in section II, their proofs in section III.
	The remainder of this introduction relates our work to the literature on the subject.

	First of all, $(i)$, $(ii)$, $(iii)$,  and $(iv)$ only express what everyone should expect to be 
true anyhow; in fact, $(i)$ and $(ii)$ are not even new.
	Namely, except for the monotonic increase, $(i)$ has been proved 
in \cite{BenguriaLieb}; the monotonic increase was proved, for $\lambda=1$, in \cite{Hog}.
	As for $(ii)$, this was shown for $\lambda \geq \lambda_*$ in \cite{BenguriaBrezisLieb};
the numerical value $\lambda_*\approx 1/1.21$ is due to \cite{Baumgartner} and \cite{Solovej}.
	Yet, our proof of $(i)$ and $(ii)$ overall is novel (to the best of the authors knowledge).

	As to $(iii)$, this result seems to be new, though
superficially it is reminiscent of the results proved recently in
\cite{BGM}, \cite{ErdosYau} (see also \cite{BEGMY}), \cite{ElgardSchlein}, \cite{FroehlichKnowlesSchwarz}, and 
in \cite{Pickl}, which establish the asymptotic exactness of the dynamical Hartree equations for Coulomb or Newton 
interactions in the limit $N\to\infty$; see also \cite{HEPP} and \cite{SPOHN} for earlier results 
with more regular interactions.
	In these works it is shown that  $\psi^{(\powN)}_t \to \phi_t^{\otimes\Nset}$ as $N\to\infty$ \emph{if}
$\psi^{(\powN)}_0 \to \phi_0^{\otimes\Nset}$ as $N\to\infty$, where
$\psi^{(\powN)}_t$ denotes the $N$-body Schr\"odinger wave function at time $t$ and 
$\psi^{(\powN)}_0$ its initial state, while $\phi_t$ is the solution to the Hartree equation at time $t$
launched by initial data $\phi_0$.
	These  ``propagation of factorization'' results are very much in the spirit of what Kac, McKean and
Lanford called ``propagation of chaos''  in their approaches to derive Boltzmann's equation in the kinetic theory 
of a dilute classical gas.
	Note that the factorization of the initial wave function (as $N\to\infty$) is being \emph{assumed} 
in these works; in fact, most $N$-body wave functions do not factorize in the limit $N\to\infty$.

	Also $(iv)$ is new.
	Previously, in \cite{BenguriaLieb} and \cite{LiebYau} (taking the $c\to\infty$ 
corollary), it was shown that 
$\int |\psi^{(\powN)}_\lambda|^2(q^{}_1,...,q^{}_N)\dd^3q^{}_2\cdots \dd^3q^{}_N\to |\phi_\lambda(q^{}_1)|^2$.
	Our $(iii)$ generalizes this to 
$\int |\psi^{(\powN)}_\lambda|^2(q^{}_1,...,q^{}_N)\dd^3q^{}_{n+1}\cdots \dd^3q^{}_N\to \prod_{k=1}^n|\phi_\lambda(q^{}_k)|^2$
for any $n\in\Nset$.
	Such results state that the \emph{expected value} of the normalized empirical $n$-particle density agrees 
with $(|\phi_\lambda|^2)^{\otimes n}$ in the limit $N\to\infty$. 
	The gap between such an average result and $(iv)$ is bridged by a law~of~large~numbers.

	Our last statement makes it plain that our approach will be probabilistic
(or: statistical-mechanical) in nature. 
	More precisely, it resembles the construction of the thermodynamic mean-field limit of the classical 
ensembles with thermodynamically unstable interactions, except that here the ensemble is Born's,
with $|\psi^{(\powN)}_\lambda|^2$ as ensemble probability density function on $\Rset^{3N}$.
	The proof exploits the fact that the finite-$N$ variational principle for $|\psi^{(\powN)}_\lambda|^2$ 
resembles the free-energy variational principle for the classical petit-canonical configurational measure, 
with the Fisher functional in the variational principle for Born's ensemble now playing the role of
the negative of the Gibbs entropy in the classical petit-canonical configurational free-energy.
	Readers familiar with the works \cite{MesserSpohn}, \cite{KieRMP}, \cite{KieJSTAT} on the 
petit- and micro-canonical classical ensembles will recognize the strategy; see also \cite{APT} for a
petit--canonical quantum ensemble which uses the classical Gibbs entropy for distributions
on Feynman--Kac path configuration space.

        We end this introduction on a cautionary note. 
        The ratio of the $\alpha$-bosonic ``electron's'' mass $m$ versus the mass $M$ of any regular nucleus is
not really so small that one could have much confidence in the Born--Oppenheimer approximation ($M\to\infty$). 
	The correct Galilei-invariant atomic (or ionic) model with a dynamical nucleus 
of mass $M$ is a more subtle $N+1$ body problem.
	Slightly simpler than this is the Galilei-invariant Newtonian gravitational analog of the problem, 
\cite{PostB}, \cite{HallD}.
Its pseudo-relativistic version, known as a ``bosonic neutron star,'' has recently attracted lots of attention,
see \cite{LiebYau},  \cite{ElgardSchlein},  \cite{HallLuchaA}, \cite{FroehlichLenzmann}, \cite{HallLuchaB}, 
\cite{FrankLenzmann}, \cite{FroehlichKnowlesSchwarz}, and \cite{Pickl}. 
        We hope to address such Galilei-invariant systems in a follow-up work.
\newpage

\section{Statement of results}
\vskip-0.3truecm
\noindent
        In this section we give a precise formulation of the $N$-body variational problem (subsection 2.1) 
and of our Hartree limit theorems (subsections 2.2 and 2.3). 
	The bosonic ions are treated in Born--Oppenheimer approximation, for simplicity.
	Also, all particles are treated as pointlike, which can only be a rough approximation
for $\alpha$-bosonic atoms and ions.

\vskip-1truecm
\subsection{The $N$-body variational principle}\label{sec:NbodyVP}
\vskip-0.2truecm
\noindent
	The Hamiltonian of a non-relativistic atomic ion with an infinitely massive nucleus 
(the Born--Oppenheimer approximation; indicated by the label $_\BO$) is the formal Schr\"odinger operator 
\begin{equation} 
H^{(\powZ,\powN)}_{\BO}
=\label{HamiltonianATOM}
\textstyle{\sum\limits_{\sss 1\leq k \leq N}^{} \left({{\frac{1}{2m}}} |\pV_k|^2 -  Zz^2e^2{{\frac{1}{|\qV_k|}}}\right)
+
\sum\sum\limits_{\hskip-.6truecm \sss 1\leq j<k\leq N}^{} z^2e^2{{\frac{1}{|\qV_j-\qV_k|}}}},
\end{equation}	
with the understanding that $\sum\sum_{1\leq j<k\leq 1}=0$.
	In \Ref{HamiltonianATOM}, $\pV_k=-i\hbar\nabla_k$ is the familiar momentum operator 
canonically dual to the $k$-th ``electron's'' configuration space position operator, $\qV_k\in\Rset^3$.
	Moreover, $m$ is the ``Newtonian inertial mass'' and $ze$ the electric charge of each of the $N\in\Nset$ 
``electrons,'' with $z=-2$ for anti-$\alpha$ particles and $z=-1$ for true fermionic electrons.
	Lastly, $Z|z|e$ (with $|z|Z\in\Nset$) is the charge of the atomic nucleus (which is fixed at the origin);
of course, $e(>0)$ is the conventional elementary electric charge.
	Although $|z|Z\in\Nset$ in nature,  we will conveniently allow $Z\in\Rset_+$.

	The formal operator $H^{(\powZ,\powN)}_{\BO}$ is densely defined on 
$(\Csp^\infty_0\cap \Lsp^2)(\Rset^{3\PowN})$.
	As self-adjoint extension we take its Friedrichs extension \cite{ReedSimonI}, 
also denoted by $H^{(\powZ,\powN)}_{\BO}$,
which is a permutation-symmetric, self-adjoint operator with form domain given by the $N$-fold tensor product 
$D_\funcQ^{(\powN)}\equiv \Hsp^1(\Rset^3)\otimes \cdots\otimes \Hsp^1(\Rset^3)\subset\Lsp^2(\Rset^{3\PowN})$.
	The quadratic form associated to the Friedrichs extension $H^{(\powZ,\powN)}_{\BO}$ is 
\begin{equation}
	\funcQ^{(\powZ,\powN)}_{\;\BO}(\Psi^{(\powN)}) 
=\label{QformCOUL}
	{\textstyle{\frac{\hbar^2}{2m}}} \funcK^{(\powN)}(\Psi^{(\powN)}) 
	-Zz^2e^2\,\funcC^{(\powN)}(\Psi^{(\powN)}) +z^2e^2\,\funcI^{(\powN)}(\Psi^{(\powN)}),
\end{equation}	

\vskip-0.1truecm
\noindent
where 

\vskip-.85truecm
\begin{eqnarray} 
	\funcK^{(\powN)}(\Psi^{(\powN)})
\hskip-0.6truecm&& =\label{Kform}
\int  {\textstyle\sum\limits_{\sss 1\leq k\leq N}^{}} |\nabla_k\Psi^{(\powN)}|^2 \dd^{^{3\powN}}\!\!\!\qV,
\\
\funcC^{(\powN)}(\Psi^{(\powN)}) 
\hskip-0.6truecm&& =
 \int {\textstyle\sum\limits_{\sss 1\leq k\leq N}^{} {\frac{1}{|\qV_k|}}}|\Psi^{(\powN)}|^2  \dd^{^{3\powN}}\!\!\!\qV,
\label{Cform}\\
\funcI^{(\powN)}(\Psi^{(\powN)}) 
\hskip-0.6truecm&& =
 \int {\textstyle\sum\sum\limits_{\hskip-.6truecm \sss 1 \leq k < l \leq N}^{}
\frac{1}{|\qV_k-\qV_l|}} |\Psi^{(\powN)}|^2  \dd^{^{3\powN}}\!\!\!\qV.
\label{Iform}
\end{eqnarray}

	The bosonic \emph{ground state energy} of $H^{(\powZ,\powN)}_{\BO}$ is defined by
\begin{equation} 
	\cE_{\BO}^{\Bose}(Z,N) 
:=\label{qmEnullCOULOMB}
	\inf\left\{\funcQ^{(\powZ,\powN)}_{\;\BO}(\Psi^{(\powN)}) \, \big| \,
	\Psi^{(\powN)}\in D_\funcQ^{(\powN)} \;;\; \|\Psi^{(\powN)}\|_{\Lsp^2(\Rset^{3N})}=1 \right\}.
\end{equation}	
	The infinimum is always finite, yet depending on the ratio of $Z$ vs. $N$ it may or may not be achieved 
by a minimizing $N$-body wave function $\Psi^{(\powZ,\powN)}_\Bose\in D_\funcQ^{(\powN)}$, called \emph{the ground state}.
	Thus, the ``ground state energy'' may or may not be the ``energy of a ground state.''

	To sort this out we recall that 
$\cE_{\BO}^{\Bose}(Z,N)\! \leq \min\sigma_{\rm ess}\big( H^{(\powZ,\powN)}_{\BO}\big)\!,$
by the min-max principle (Theorem XIII.1 in \cite{ReedSimonIV}).
	So either $\cE_{\BO}^{\Bose}(Z,N) = \min\sigma_{\rm ess}\big( H^{(\powZ,\powN)}_{\BO}\big)$ or 
$\cE_{\BO}^{\Bose}(Z,N) < \min\sigma_{\rm ess}\big( H^{(\powZ,\powN)}_{\BO}\big)$.

	When $\cE_{\BO}^{\Bose}(Z,N) = \min\sigma_{\rm ess}\big( H^{(\powZ,\powN)}_{\BO}\big)$, then 
a minimizer $\Psi^{(\powZ,\powN)}_\Bose$ may or may not exist, indeed (to the best of the authors knowledge). 
	However, the bottom of the essential spectrum of a general self-adjoint operator can only be one
or more of the following:
(a) an eigenvalue of infinite multiplicity (e.g. the eigenvalue 0 of a projection operator onto a 
finite-dimensional subspace of Hilbert space), 
(b) a limit point of the pure point spectrum (as is the case for $H^{(\powZ,\sss{1})}_{\BO}$), or 
(c) the bottom of the continuous spectrum (as is the case for $H^{(\powZ,\sss{1})}_{\BO}$ also; 
note that the possibilities are not mutually exclusive). 
	Clearly, in case (a) the ground state would be infinitely degenerate, and if a
ground state exists in cases (b) or (c), even if different from case (a), it would be ``as close as it gets'' to 
being infinitely degenerate without literally being infinitely degenerate; we will call this 
\emph{infinitely quasi-degenerate}.
	Paraphrasing Simon  \cite{SimonHPA}, all such ground states ``are somewhat pathological beasts.''

	On the other hand, when $\cE_{\BO}^{\Bose}(Z,N)< \min\sigma_{\rm ess}\big( H^{(\powZ,\powN)}_{\BO}\big)$, 
the existence of a minimizing $\Psi^{(\powZ,\powN)}_\Bose$ is guaranteed by the min-max principle.
	In this case $\Psi^{(\powZ,\powN)}_\Bose$ is in fact nondegenerate and it can be taken to be 
real-valued and positive.
	Moreover, the permutation symmetry of $H^{(\powZ,\powN)}_{\BO}$ implies that
$\Psi^{(\powZ,\powN)}_\Bose$ is automatically permutation symmetric, too, i.e. bosonic.
	All this is of course well-known; see, e.g., \cite{Lieb}, \cite{LiebSeiringer}.

	For our approach we will need the existence of a nondegenerate $\Psi^{(\powZ,\powN)}_\Bose\in D_\funcQ^{(\powN)}$.
	So we follow \cite{SimonHPA} and consider $\Psi^{(\powZ,\powN)}_\Bose$ to be \emph{a proper ground state} if and 
only if $\cE_{\BO}^{\Bose}(Z,N)$ \emph{belongs to the discrete spectrum}, equivalently 
$\cE_{\BO}^{\Bose}(Z,N) < \min\sigma_{\rm ess}\big( H^{(\powZ,\powN)}_{\BO}\big)$.
	Henceforth $\Psi^{(\powZ,\powN)}_\Bose$ always denotes such a proper ground state.

	The condition
$\cE_{\BO}^{\Bose}(Z,N) < \min\sigma_{\rm ess}\big( H^{(\powZ,\powN)}_{\BO}\big)$ can be 
expressed in a more user-friendly manner.
	This is accomplished by the HVZ theorem, which asserts that 
$\sigma_{\rm ess}\big( H^{(\powZ,\powN)}_{\BO}\big)=[\cE_{\BO}^{\Bose}(Z,N-1) ,\infty)$, see
Theorem 11.2 in \cite{Teschl} (this is a special case of the HVZ theorem for $ H^{(\powZ,\powN)}_{\sss\mathrm{M}}$ 
with nuclear mass $M<\infty$, see Theorem XIII.17  in \cite{ReedSimonIV}).
	So a proper ground state  $\Psi^{(\powZ,\powN)}_\Bose$ exists
whenever $\cE_{\BO}^{\Bose}(Z,N) < \cE_{\BO}^{\Bose}(Z,N-1)$. 
	We need to know for which combinations of $Z$ and $N$ this inequality is fulfilled.

\subsubsection{The nuclear charge $Z$ necessary to bind $N$ bosonic electrons}\label{sec:Zstar}

        By Zhislin's theorem (Theorem 12.2 in \cite{LiebSeiringer}), a proper $N$-body ground state 
$\Psi^{(\powZ,\powN)}_\Bose$ exists whenever $Z>N-1$.
	On the other hand, Benguria and Lieb \cite{BenguriaLieb} have shown that 
$\Psi^{(\powZ,\powN)}_\Bose$ exists whenever $Z\geq \lambda_* N$ \emph{provided $N$ is large enough},
with $\lambda_*^{}=1/(1+\gamma)$ in the notation of  \cite{BenguriaLieb},
and numerically $\gamma\approx 0.21$ \cite{Baumgartner}.
	Combining these two results we obtain the following.
\begin{cor}
\label{coro:lambdaSTAR}
	There exists a $\lambda^*$ satisfying $1> \lambda^* \geq \lambda_*$ such that for all $N\in\Nset$
a proper ground state wave function $\Psi^{(\scriptscriptstyle{\lambda N},\powN)}_\Bose$ for \Ref{qmEnullCOULOMB} exists 
whenever $\lambda >\lambda^*$, and there exists some $N^*\in\Nset$ such that for all $N\geq N^*$
a proper ground state wave function $\Psi^{(\scriptscriptstyle{\lambda N},\powN)}_\Bose$ for \Ref{qmEnullCOULOMB} exists 
whenever $\lambda \geq \lambda_*\approx 1/1.21$.
\end{cor}
\begin{rem}
\hskip-2pt
	In view of Corollary \ref{coro:lambdaSTAR} we will have to restrict $N$ to be ``big enough''
in the proofs of our main theorems, which are stated for the optimal parameter region $\lambda\geq \lambda_*$.
	Whenever $\lambda>\lambda^*$, this restriction on $N$ can be dropped; 
in particular, this is the case if $\lambda\geq 1$.
\end{rem}

        We emphasize that any nontrivial values of $\lambda^*$ and $N^*$, i.e. values
other than $\lambda^*=\lambda_*$ and $N^*=1$, whatever they might be, will presumably just be
artifacts of our incomplete knowledge about the full range 
of $Z$ values, given $N$, for which a proper bosonic ground state exists.
        Indeed, as made plainly clear by the results of \cite{BenguriaLieb}, Zhislin's theorem does not 
reveal the full range  of such $Z$ values. 
        We next state a conjecture as to what this range could be.

        We define $Z_*(N)$, the minimum amount of nuclear charge needed 
(in the sense of: to be exceeded) to properly bind $N$ ``bosonic electrons,'' thus:
\begin{equation} 
\hskip-5pt
	Z_*(N)
:=\label{Lstar}
	\inf\big\{Z\!\in\Rset_+\big| \exists \Psi^{(\powZ,\powN)}_\Bose\!\! \in\! D_\funcQ^{(\powN)}
	 s.t.\, \funcQ^{(\powZ,\powN)}_{\;\BO}(\Psi^{(\powZ,\powN)}_\Bose) < \cE_{\BO}^{\Bose}(Z,N-1)\! \big\}.
\end{equation}	
\begin{rem}
	A ground state may or may not exist when $Z= Z_*(N)$; however, such a ground state would belong to the 
essential spectrum and, therefore, not be proper in the sense stipulated above.
	Incidentally, it is even conceivable that non-proper ground states exist for $Z<Z_*(N)$. 
\end{rem}
	By Zhislin's theorem, and the virial theorem, we have $0\leq Z_*(N)\leq N - 1$ for all $N\in\Nset$; in 
particular, $Z_*(1)=0$ (the hydrogenic problem), and  $Z_*(2)\leq 1$ (the Helium-type problem; discussed 
extensively in section 4.3 of \cite{Thirring}).
        Also recall that Zhislin's theorem states that a proper ground state 
$\Psi^{(\powZ,\powN)}_\Bose$ exists whenever $Z>N-1$.
        Thus it is suggestive to suspect the following.
\begin{con}
\label{conj:minZperN}
	A proper $N$-body ground state wave function $\Psi^{(\powZ,\powN)}_\Bose$ for \Ref{qmEnullCOULOMB} exists 
whenever $Z>Z_*(N)$. 
\end{con}
\begin{rem}
        Conjecture \ref{conj:minZperN} is the analogue of a claim in \cite{ReedSimonIV}, made
(without proof) after their proof of Theorem XII.9 in their discussion of ``example 1 revisited.''
	Indeed, just replace their $H_0\, (\equiv -\Delta)$ by our $ H^{(\sss{0},\powN)}_{\BO}$.
\end{rem}

	We note that the function $N\mapsto Z_*(N)$ is a cousin of Lieb's and Benguria's $Z\mapsto N_{\mathrm{max}}(Z)$, 
which is the maximum number of (here: bosonic) electrons which a nucleus of charge $Z$ can bind. 
        Interestingly enough,  the precise analogue of Conjecture \ref{conj:minZperN} for $Z\mapsto N_{\mathrm{max}}(Z)$
is also an open problem.
\begin{con}
\label{conj:ZbindsNandNminONE}
	A proper $N$-body ground state wave function $\Psi^{(\powZ,\powN)}_\Bose$ 
for \Ref{qmEnullCOULOMB} exists for all $1\leq N\leq N_{\mathrm{max}}(Z)$.
\end{con}
\begin{rem}
	Conjecture \ref{conj:ZbindsNandNminONE} is phrased in \cite{LiebSeiringer} more generally thus: 
\emph{If a nucleus with $Z$ charges binds $N$ electrons (and $N\geq 2$), then the 
same nucleus also binds $N-1$ electrons.}
        This seemingly obvious ``truth'' has only been proven to be true for $N< Z+2$, by Zhislin's theorem,
for both bosonic and fermionic electrons.
	I thank the referee for drawing my attention to this conjecture, which 
is mentioned as an open problem in Chpt.12 of \cite{LiebSeiringer}.
\end{rem}
	To state our last conjecture we define $\Lambda_*(N):=Z_*(N)/N$. 
        It follows from the work in \cite{BenguriaLieb} and \cite{Solovej} that 
$\lim_{N\uparrow\infty}\Lambda_*(N)=\lambda_*\approx 1/1.21$.
        Moreover, $\Lambda_*(1)=0$ and $0<\Lambda_*(2)\leq 1/2$.
        The upper estimate on $\Lambda_*(2)$ follows from Zhislin's theorem, which yields the monotone 
increasing upper bound $\Lambda_*(N)\leq 1 - \frac1N$ for all $N\in\Nset$.
	We suspect that $\Lambda_*(N)$ itself is monotone.
\begin{con}
\label{conj:monoUPz}
        The map $N\mapsto \Lambda_*(N)$ is monotonic increasing.
\end{con}
\begin{rem}
	Conjectures \ref{conj:minZperN} and \ref{conj:monoUPz}, if true, would imply that 
$N^*=1$, so that in the proofs of our main theorems we could drop the restriction $N\geq N^*$.
\end{rem}

        We now move on to state our Hartree limit theorems.



\subsection{The Hartree approximation}

	The Hartree approximation to the bosonic ground state energy $\cE_{\BO}^{\Bose}(Z,N)$
is obtained by estimating $\funcQ^{(\powZ,\powN)}_{\;\BO}(\Psi^{(\powZ,\powN)}_\Bose)$ from above with the help of 
convenient trial wave functions $\Psi^{(\powN)}\equiv\Phi^{\otimes \powN}$, 
with $\Phi\in\Hsp^1(\Rset^3)$ satisfying $\|\Phi\|_{\Lsp^2(\Rset^{3})}=1$.
	We thus have 
$\cE_{\BO}^{\Bose}(Z,N)\leq \inf_\Phi \funcQ^{(\powZ,\powN)}_{\;\BO}(\Phi^{\otimes\powN})
			=   \inf_\Phi \cH^{(\powZ,\powN)}_{\BO}(\Phi)$, 
where
\begin{equation}	
\hskip-5pt
	\cH^{(\powZ,\powN)}_{\BO}(\Phi) 
=\label{HartreeCoulN}
	N{\textstyle{\frac{\hbar^2}{2m}}} \funcK^{(1)}(\Phi) -NZz^2e^2\, \funcC^{(1)}(\Phi) +
	N(N-1)z^2e^2\tfrhalf\, \funcI^{(2)}(\Phi\otimes\Phi)
\end{equation}	
is the Hartree functional for $H^{(\powZ,\powN)}_{\BO}$.

	We are interested in the large-$N$ behavior, with scaling $Z=\lambda N$.
	Setting $\Phi(\qV)=N^{3/2}\phi(N\qV)$ in \Ref{HartreeCoulN} 
yields $\cH^{(\lambda\powN,\powN)}_{\BO}(\Phi) =N^3\funcH_{\BO,\lambda}(\phi)+O(N^2)$,~where
\begin{equation}	
	\funcH_{\BO,\lambda}(\phi) 
=\label{HartreeCoul}
	{\textstyle{\frac{\hbar^2}{2m}}} \funcK^{(1)}(\phi) - \lambda z^2e^2\, \funcC^{(1)}(\phi) +
	z^2e^2\tfrhalf\, \funcI^{(2)}(\phi\otimes\phi)
\end{equation}	
is the ``asymptotic Hartree functional,'' and $O(N^2) = - N^2 z^2e^2\tfrhalf\, \funcI^{(2)}(\phi\otimes\phi)$
a negative correction term.
	Dropping this negative correction term results in the  \emph{uniform} (in $N$) upper bound 
$N^{-3} \cE_{\BO}^{\Bose}(\lambda N,N)\leq \inf_\phi \funcH_{\BO,\lambda}(\phi)\;\forall\; N$, where the
infimum is taken over $\phi\in\Hsp^1(\Rset^3)$ satisfying $\|\Phi\|_{\Lsp^2(\Rset^{3})}=1$.
	Although this bound on $N^{-3} \cE_{\BO}^{\Bose}(\lambda N,N)$ is weaker than the bound
$N^{-3} \cE_{\BO}^{\Bose}(\lambda N,N)\leq N^{-3} \inf_\Phi \cH^{(\lambda\powN,\powN)}_{\BO}(\Phi)$, 
the omission of the correction term worsens this upper bound only by an amount of relative order $N^{-1}$, 
which vanishes as $N\to\infty$.

	The Hartree approximation to the bosonic ground state energy, indeed to the bosonic
ground state itself, becomes exact as $N\to\infty$, provided $\lambda\geq \lambda_*:=\lim \Lambda_*(N)$.
	This is made precise by the following theorem.
\begin{thm}
\label{thm:HartreeC} (Asymptotic exactness of Hartree theory for bosonic ions.)

\noindent
	Whenever $\lambda\geq \lambda_*$, then

\hskip7pt $(i)$ $N^{-3}\cE^{\Bose}_{\BO}(\lambda N,N) \nearrow \inf\{\funcH_{\BO,\lambda}(\phi):
	\phi\in \Hsp^1\,\&\, \|\phi\|_{\Lsp^2}=1\}$ as $N\to\infty$.

\noindent
	Moreover, 

\hskip3.5pt $(ii)$ $\funcH_{\BO,\lambda}(\phi)$ has a unique (positive) minimizer, 
$\phi_{\lambda}\in \Hsp^1 \cap \{\|\phi\|_{\Lsp^2}=1\}$,

\noindent
and then, setting 
$
\psi^{(\powN)}_\lambda(q^{}_1,...,q^{}_N)
:=
N^{-\sfrTHREEhalf N}\Psi^{(\lambda\powN,\powN)}_\Bose(N^{-1}q^{}_1,...,N^{-1}q^{}_N)
$,
we have 

	$(iii)$ $\psi^{(\powN)}_\lambda \withNto \phi_{\lambda}^{\otimes\Nset}$, 

\noindent
in the sense that, for any $n\in\Nset$, weakly in $\Lsp^1\cap\Lsp^{3n/(3n-2)}$ we have,  

	$(iii)^\prime$ $\int |\psi^{(\powN)}_\lambda|^2(q^{}_1,...,q^{}_N)\dd^3q^{}_{n+1}\cdots \dd^3q^{}_N\withNto 
			\prod\limits_{\sss 1\leq k\leq n}^{} |\phi_{\lambda}(q^{}_k)|^2$.
\end{thm}
\begin{rem}
	Except for the monotonicity, item $(i)$ has previously been proved in \cite{BenguriaLieb}, 
and so has $(iii)^\prime$ in the special case $n=1$; 
and item $(ii)$ has previously been proved in \cite{BenguriaBrezisLieb}.
	As stated for all $n\in\Nset$, item $(iii)$, viz. $(iii)^\prime$, is novel.
	Also, the proof of Theorem \ref{thm:HartreeC} overall seems to be novel.
	It is an ``information variation on the entropy theme played'' in 
\cite{MesserSpohn}, \cite{CLMP}, \cite{KieCPAM}, \cite{APT}, \cite{ChaKieDMJ}, \cite{KieRMP}, \cite{KieJSTAT},
\cite{KieWang}.
\end{rem}	
\begin{rem}
	The monotonic increase of $N\mapsto N^{-3}\cE^{\Bose}_{\BO}(\lambda N,N)$ is a novel result, too,
strictly speaking; however, Hogreve \cite{Hog} already proved for neutral Born--Oppenheimer atoms that 
$\cE_{\BO}^{\Bose}(N,N)/N^3$ grows monotonically in $N\geq 1$, and inspection of Hogreve's proof, see his 
formulas (7)--(10), reveals that it generalizes verbatim to Born--Oppenheimer ions, without restrictions on $\lambda$.

	A slightly weaker atomic result, that $\cE_{\BO}^{\Bose}(N,N)/N^2(N-1)$ grows monotonically in $N\geq 2$,
was earlier proved in \cite{KieJSPb}.
	Also this result generalizes to ions, with proper restrictions on $\lambda$ in place,
i.e. for $\lambda>\lambda^*$ the map $N\mapsto \cE_{\BO}^{\Bose}(\lambda N,N)/N^2(N-1)$ 
is monotonic increasing.
	Of course we are happier with Hogreve's result; also, his proof, which uses
the bosonic symmetry of the wave functions, is simpler than the one in \cite{KieJSPb}.
	On the other hand, the proof in \cite{KieJSPb}, which does not use the bosonic symmetry of the ground
state wave functions, implies also that $\cE_{\BO}^{\Ferm}(N,N)/N^2(N-1)$ grows monotonically in $N\geq 2$,
where 
\begin{equation} 
	\cE_{\BO}^{\Ferm}(N,N) 
:=
	\min\left\{\funcQ^{(\powN)}_{\;\BO}(\Psi^{(\powN)}) \, \Big| \,
	\Psi^{(\powN)}\in \hat{D}_\funcQ^{(\powN)}
	 \;;\; \|\Psi^{(\powN)}\|_{\Lsp^2(\Rset^{3\powN})}=1 \right\},
\label{qmEnullCOULOMBfermi}
\end{equation}	
with $\hat{D}_\funcQ^{(\powN)}\!\equiv \Hsp^1(\Rset^3)\wedge \cdots\wedge \Hsp^1(\Rset^3)$,
is the fermionic ground state energy of $H^{(\powN,\powN)}_{\BO}\!.$ 
	Since the fermionic ground state $\Psi^{(\powN)}_{\Ferm}$ for \Ref{qmEnullCOULOMBfermi}
is anti-symmetric by construction, it is not clear whether $\cE_{\BO}^{\Ferm}(N,N)/N^3$ grows monotonically in $N$.
\end{rem}

\vskip-1truecm
$\phantom{nix}$
\subsection{Born's statistical ensemble}\label{sec:BornEns}
\vskip-0.2truecm
\noindent
	To state our next theorem, we recall Born's statistical interpretation of Schr\"o\-dinger wave functions.
	According to Born, $|\psi^{(\powN)}|^2(q^{}_1,...,q^{}_N)$ is the joint $N$-body probability density for 
finding particle 1 at $q^{}_1$, particle 2 at $q^{}_2$, and so on, when one performs an ideal measurement that 
yields the positions of the ``electrons'' in state $\psi^{(\powN)}$.
	More precisely, let $Q^{}_k$ be the random position which shows up in an ideal 
position measurement of particle $k$, let $Q^{(\powN)}\in\Rset^{3\PowN}$ be the random vector whose $k$-th 
component triple is $Q^{}_k$, and similarly let $q^{(\powN)}\in\Rset^{3\PowN}$ be the vector value taken by
$Q^{(\powN)}$, then for the ion's bosonic ground state,

\vskip-15pt
\begin{equation}
	\mathrm{Prob}\left(Q^{(\powN)}\in \ddN{q}\;\mbox{\rm{near}}\, q^{(\powN)}\right)
= \label{eq:BORNprobFORMULA}
	|\psi^{(\powN)}_\lambda|^2(q^{}_1,...,q^{}_N)\textstyle\prod\limits_{\sss 1\leq k\leq N}^{}\dd^3q^{}_k.
\end{equation}

\vskip-8pt
\noindent
	For each $n\!\leq\!N$, any $Q^{(\powN)}$ maps uniquely into the empirical random $U$-statistics
\begin{equation}
	\uli\Delta^{(n)}_{Q^{(\powN)}}\!(s_1,\cdots\!,s_n)
 = \label{normalEMPmeasUn}
	{\textstyle{\big(\genfrac{}{}{0pt}{}{N}{n}{}{}\big)^{-1}\;
\sum\cdots\hskip-.4truecm\sum\limits_{\hskip-.85truecm \sss 1\leq k_1 < \cdots <k_n\leq N}^{}}
\, \prod\limits_{\sss 1\leq j\leq n}}\delta_{Q^{}_{_{k_j}}}\!\!(s_j),
\end{equation} 
where the $s_j\in\Rset^3$ are generic points in physical space $\Rset^3$; see \cite{EichelsbacherSchmock}.
	If one factors out the permutation group $S_N$, 
then the mapping between $Q^{(\powN)}$ and any of its empirical $U$-statistics is actually one-to-one.
	These empirical $U$-statistics are themselves random measures (normalized counting measures of random point 
configurations), and the first few of them have practical significance. 
	In particular, $\uli\Delta^{(1)}_{q^{(\powN)}}(s)$ is the normalized ``electron'' density of the bosonic atom
found in an ideal density measurement, or rather as computed from an ideal measurement of all  positions 
of the bosonic ``electrons.''
	For finite $N$ the empirical electron density will fluctuate from any one such ideal measurement to
the next, but fluctuations subside when $N\to\infty$.
	More precisely, the following \emph{law-of-large-numbers}-type result is proved in section 3.2:
\begin{thm}
\label{thm:HartreeLIMITrho} (Asymptotic absence of density fluctuations.)

\noindent
For $\lambda\geq \lambda_*$, and for any $n\in\Nset$, when $N\to\infty$, then 

\hskip1pt $(iv)$ $\uli\Delta^{(n)}_{Q^{(\powN)}}\to |\phi_\lambda|^2\otimes \cdots\otimes|\phi_\lambda|^2\ 
(n$-fold product),

\noindent
in the sense that when ``dist'' means Kantorovich--Rubinstein distance, then

\hskip1pt $(iv)^\prime$  
$\mbox{\rm{Prob}}
	\left(\mbox{\rm{dist}}\left(\uli\Delta^{(n)}_{Q{(\powN)}}, (|\phi_{\lambda}|^2)^{\otimes n}\right)>\eps\right)
		\to 0\,	\forall\,\eps>0$.
\end{thm}

\begin{rem}
	The case $n=1$ in Theorem \ref{thm:HartreeLIMITrho} supplies the proper meaning for the notion of
$N|\phi_\lambda|^2$ as (continuum approximation of) the ``particle density.''
\end{rem}

\section{Proofs of the Theorems}

\subsection{Proof of Theorem \ref{thm:HartreeC} }
	Our strategy is to adapt the construction of the thermodynamic mean-field limit of 
the classical petit-canonical configurational ensemble with thermodynamically unstable 
interactions\footnote{See \cite{MesserSpohn} for the original paper, where the classical petit-canonical 
	configurational ensemble is treated for particles with Lipschitz-continuous interactions; see
	also \cite{KieRMP}, \cite{KieJSTAT} for the classical micro-canonical ensemble with singular 
	interactions.}
to ``Born's ground state  ensemble.''
	The proofs exploit the fact that the finite-$N$ variational principle for Born's ensemble
resembles the free-energy variational principle for the classical petit-canonical configurational measure, 
with the Fisher functional in the variational principle for Born's ensemble measure now playing the role of
the negative of the Gibbs entropy in the free-energy functional for the classical petit-canonical configurational 
ensemble measure.

	We give the proofs for bosonic Born--Oppenheimer ions in general.
	The neutral atoms with $Z=N$ are included as special case. 
%
\subsubsection{Ensemble reformulation of Theorem \ref{thm:HartreeC} }

	The proof of Theorem \ref{thm:HartreeC}  uses Born's statistical ensemble interpretation of 
the rescaled ground state wave function $\psi^{(\powN)}_\lambda$ of the Hamiltonian \Ref{HamiltonianATOM} 
only to the extent that $\int|\psi^{(\powN)}_\lambda|^2\,\ddN{q} =1$, so that \emph{formally} the
permutation-symmetric $|\psi^{(\powN)}_\lambda|^2$ can be ``called'' an ensemble probability density.
	Truly probabilistic arguments will be employed eventually in the proof of Theorem \ref{thm:HartreeLIMITrho},
but the control for Theorem \ref{thm:HartreeC}  is accomplished using functional-analytical 
and measure-theoretical arguments, only worded in statistical mechanics lingo.

	We write $|\psi^{(\powN)}_\lambda|^2=:\varrho^{(\powN)}_\lambda\in(\Psp^s\cap\Lsp^1)(\Rset^{3\PowN})$,
by which we denote the permu\-tation-symmetric probability measures on $\Rset^{3\PowN}$ which are absolutely 
continuous w.r.t. Lebesgue measure.
	More generally, any normalized and rescaled bosonic $N$-body wave function $\psi^{(\powN)}$ 
defines a formal probability measure $\varrho^{(\powN)}\in(\Psp^s\cap\Lsp^1)(\Rset^{3\PowN})$ through
$|\psi^{(\powN)}|^2=: {\varrho^{(\powN)}}$.
	Now recalling that $\int|\nabla\Psi|^2\dd q = \int|\nabla|\Psi||^2\dd q$, we see that upon
defining\footnote{For optical 
		reasons we prefer the notation $\surd{(\,\cdot\,)}$ over $\sqrt{\big.(\,\cdot\,)}$.}
$\surd{\varrho^{(\powN)}}(q^{}_1,...,q^{}_N)=
	N^{-\sfrTHREEhalf N}|\Psi^{(\powN)}|(N^{-1}q^{}_1,...,N^{-1}q^{}_N)\geq 0$, we can 
express the quadratic energy functional \Ref{QformCOUL} as a functional of $\varrho^{(\powN)}$.
	Also pulling out a factor $N^2$, we get 
$\funcQ^{(\lambda\powN,\powN)}_{\;\BO}(\Psi^{(\powN)}) = N^2 \pzcE_{\BO,\lambda}^{(\powN)}(\varrho^{(\powN)})$, with
\begin{equation}
	\pzcE_{\BO,\lambda}^{(\powN)}\!\left(\varrho^{(\powN)}\right) 
=\label{qCOULOMBenergyRESCALED}
	{\textstyle{\frac{\hbar^2}{8m}}}
	\pzcF^{(\powN)}\!\left(\varrho^{(\powN)}\right) 
	+
	z^2e^2 \pzcV_\lambda^{(\powN)}\!\left(\varrho^{(\powN)}\right) ,
\end{equation}
where, for $\varrho^{(\powN)}\in \Psp^s(\Rset^{3\PowN})$ with $\surd{\varrho^{(\powN)}}\in \Hsp^1(\Rset^{3\PowN})$, 
\begin{equation}
	\pzcF^{(\powN)}\!\left(\varrho^{(\powN)}\right) 
=\label{kinErhoNfctl}
	{4}\int\! \left|\nabla \surd{\varrho^{(\powN)}}\right|^2 \ddN{q}
\end{equation}
and
\begin{equation}
	\pzcV_\lambda^{(\powN)}\!\left(\varrho^{(\powN)}\right) 
=\label{potErhoNfctl}
	 \int\! \Big(\!-\lambda\!\!{\textstyle\sum\limits_{\sss 1\leq k\leq N}^{} {\frac{1}{|q_k|}}}
     + \tfrN \;{\textstyle\sum\sum\limits_{\hskip-.6truecm \sss 1 \leq k < l \leq N}^{} \frac{1}{|q_k-q_l|}} 
	\Big)\varrho^{(\powN)}\ddN{q}.
\end{equation}
	The functional $\pzcE^{(\powN)}_{\BO,\lambda}(\varrho^{(\powN)})$ we call 
``the quantum energy of $\varrho^{(\powN)}$.'' 
	It inherits from $\pzcQ^{(\lambda\powN,\powN)}_\BO$ the following
properties: if  $\lambda\geq \lambda^*$ and  $N\geq N^*$, the functional
\Ref{qCOULOMBenergyRESCALED} achieves its infimum at the unique and permutation-symmetric probability density
$
\varrho^{(\powN)}_\lambda = |\psi^{(\powN)}_\lambda|^2
$,
i.e.
\begin{equation}
	\inf_{\varrho^{(\powN)}\in \Psp^s(\Rset^{3N})} \pzcE_{\BO,\lambda}^{(\powN)}\bigl(\varrho^{(\powN)}\bigr) 
=\label{vpErescaled}
	\pzcE_{\BO,\lambda}^{(\powN)}\bigl(|\psi^{(\powN)}_\lambda|^2\bigr);
\end{equation}
\vskip-.1truecm
\noindent
if $\lambda\!>\!\lambda^* \in[\lambda_*,1)$, then \Ref{vpErescaled} holds $\forall\, N\!\geq\!1$.
	Moreover,
\begin{equation}
\label{vpFEzweiHOLO}
	\pzcE_{\BO,\lambda}^{(\powN)}\bigl(\varrho^{(\powN)}_\lambda\bigr) 
= 
	N^{-2}\cE^{\Bose}_{\BO}(\lambda N,N).
\end{equation}

	We next draw \emph{the parallel to classical statistical mechanics}.
	Namely, while  \Ref{potErhoNfctl} equals
the \emph{potential energy of} $\varrho^{(\powN)}$ (up to the factor $z^2e^2$),
with $\pzcV$ as usual standing for ``Volta,''\footnote{We might have alternatively
	chosen $\pzcC$, to stand for ``Coulomb,'' but we already used $\pzcC$ to denote the expected value 
	of the ``central Schr\"odinger potential.''}
and \Ref{kinErhoNfctl} equals the \emph{kinetic energy of} $\varrho^{(\powN)}$ (up to the factor $\hbar^2/8m$,
with the strange factor $1/8$ compensated for by the factor 4 in  \Ref{kinErhoNfctl}),
the rationale for the notation $\pzcF$ is that \Ref{kinErhoNfctl} is identical to 
\begin{equation}
	\pzcF^{(\powN)}\!\left(\varrho^{(\powN)}\right) 
=\label{FISHERfctlN}
	\int \left|\nabla \ln \varrho^{(\powN)}\right|^2 \varrho^{(\powN)}\ddN{q},
\end{equation}
which was introduced by R. Fisher \cite{FisherR} as measure for the information content of $\varrho^{(\powN)}$
regarding its mean. 
	This \emph{Fisher information of} $\varrho^{(\powN)}$ shares many properties with (the negative of) the
\emph{Gibbs entropy of} $\varrho^{(\powN)}$, defined  by\footnote{Normally in classical statistical mechanics
		the Gibbs entropy of a probability density function on \emph{phase space} 
		$\Rset^{3N}\times \Rset^{3N}$ is taken relative to the uniform density $h^{-3N}$, which is the 
		only place where one can make (a minimal) contact with quantum physics.
		Since we here work  on \emph{configuration space}  $\Rset^{3N}$ we do the next best thing and
		define the Gibbs entropy relative to the uniform density $(\hbar/mc)^{-3N}$, where $\hbar/mc$ 
		is the Compton wavelength of the ``bosonic electrons.''}
\begin{equation}
	-\pzcS^{(\powN)}\!\left(\varrho^{(\powN)}\right) 
=\label{GIBBSentropy}
	 \int \ln \left((\hbar/mc)^{3\powN}\varrho^{(\powN)}\right) \varrho^{(\powN)} \ddN{q}
\end{equation}
if $\varrho^{(\powN)}\ln \varrho^{(\powN)}\in \Lsp^1(\Rset^{3\PowN})$; otherwise
$\pzcS^{(\powN)}\big(\varrho^{(\powN)}\big)$ is undefined (it could be $\pm\infty$).
	Note that replacing $\pzcF$ by $-\pzcS$ in \Ref{qCOULOMBenergyRESCALED} gives us
a formal classical configurational free-energy functional, with temperature $mc^2/8$ (when lengths are
normalized to $\hbar/mc$), with $c$ the speed of light in vacuo.
	Although that free-energy functional has no infimum, for the attractive Coulomb
singularity of the central Schr\"odinger potential is catastrophic in three dimensions,\footnote{Interestingly,
		the classical Coulombic free-energy functional is non-catastrophic in two dimensions, provided 
		the temperature is higher than a critical value; 
		cf. \cite{CLMP}, \cite{KieCPAM}, \cite{KieJSTAT})}
the analogy is compelling enough to suggest an adaptation of the strategy for constructing the thermodynamic 
mean-field limit of the classical petit-canonical configurational ensemble with thermodynamically unstable but 
bounded interactions  \cite{MesserSpohn} to ``Born's ground state  ensemble'' for true Coulomb interactions.
	Indeed, we will now characterize the limit points of the sequence
$\{\varrho^{(\powN)}_\lambda\}^{}_{N\geq N^*}$ in the spirit of the approach pioneered in \cite{MesserSpohn}.


\begin{thm}
\label{thm:HARTREElimitSTATES} \vskip-5pt
(Hartree limit of Born's ground state ensemble)

\noindent
	For any $n\in \Nset$ and $\lambda\geq\lambda_*$, 
let $\varrho^{(\powN)}_{\lambda,n}$ denote the $n$-th marginal probability measure of $\varrho^{(\powN)}_{\lambda}$. 
	Then the following holds:
\medskip

\noindent
(a) weakly in $(\Psp^s\cap\Lsp^\wp)(\Rset^{3n})$ with $\wp=3n/(3n-2)$, we have
\begin{equation}
	\lim_{N\to\infty} \varrho^{(\powN)}_{\lambda,n}
= \label{limBORN}
	\rho_{\lambda}^{\otimes n};
\end{equation}

\noindent
(b) weakly in $\Hsp^1(\Rset^{3n})$, we have
\begin{equation}
	\lim_{N\to\infty} \surd{\varrho^{(\powN)}_{\lambda,n}}
= \label{limSURDrho}
	\surd{\rho_{\lambda}^{\otimes n}};
\end{equation}

\noindent
(c) $\rho_\lambda\in\Psp(\Rset^3)\cap\{ \surd{\rho}\in\Hsp^1(\Rset^3)\}$ is the unique minimizer of the 
functional 
\begin{equation}
	\frake_{\BO,\lambda}(\rho) 
:= \label{standardHARTREErhoFCTL}
	{\textstyle{\frac{\hbar^2}{8m}}}\pzcF^{(1)}(\rho) 
	-	z^2e^2 \lambda \funcC^{(1)}(\surd\rho) 
	+ z^2e^2 \tfrhalf\, \funcI^{(2)}(\surd\rho\otimes\surd\rho)\big);
\end{equation}

\noindent
(d) moreover, 
\begin{equation}
\frake_{\BO,\lambda}(\rho_\lambda) = \veps_\infty (\lambda)\qquad \forall\lambda\geq\lambda_*,
\end{equation}
where
\begin{equation}
	\veps_\infty(\lambda)
:= \label{vepsLIMlambda}
	\lim_{N\to\infty}N^{-3}\cE^{\Bose}_{\BO}(\lambda N,N)
\end{equation}
is a negative, strictly monotone decreasing, concave function of $\lambda>\lambda_*$.
\end{thm}
\begin{rem}
	The functional $\frake_{\BO,\lambda}(\rho)$ defined in \Ref{standardHARTREErhoFCTL} is just 
a rewriting of the asymptotic Hartree functional $\funcH_{\BO,\lambda}(\phi)$, defined in \Ref{HartreeCoul},
in terms of $\surd\rho = \phi$.
	Therefore its unique minimizer $\rho_\lambda^{}=|\phi_\lambda^{}|^2$, where $\phi_\lambda^{}$ minimizes
the asymptotic Hartree functional.
\end{rem}
\smallskip\noindent
\subsubsection{Proof of Theorem \ref{thm:HARTREElimitSTATES}}

\vskip-2.4pt
\noindent
	We begin by stating an auxiliary result which proves item $(i)$ in Theorem \ref{thm:HartreeC}.
\vskip-2pt
\begin{prp}
\label{prop:EmonoUPion}
        The map 
$N\mapsto \inf_{\varrho^{(\powN)}}{\textstyle\frac{1}{N}} \pzcE_{\BO,\lambda}^{(\powN)}\bigl(\varrho^{(\powN)}\bigr)$ 
is monotonic increasing.
\end{prp}
\smallskip\noindent
{\textit{Proof of Proposition \ref{prop:EmonoUPion}:}} 

\noindent
	We exploit the permutation symmetry of the problem in the manner
pioneered in \cite{PostA}, \cite{PostB}, and advanced in \cite{HallB}, \cite{HallC}, \cite{Hog}.
	For $N\geq 2$ the Hamiltonian \Ref{HamiltonianATOM} is a permutation-symmetric sum of 
one- and two-body operators, and to find the infimum of its spectrum it suffices to let it act 
on the closed subdomain of bosonic (i.e. permutation-symmetric) $N$-body wave functions $\Psi^{(\powN)}$.
	This allows one to rewrite the energy functional \Ref{QformCOUL} as a conditional $\Psi^{(\powN)}$-average 
of a two-body operator for all $N\geq 2$. 
	This transforms the bosonic ground state problem of the Hamiltonian \Ref{HamiltonianATOM}
into the realm of Helium-type ground state problems.
	We implement a variation of this theme.
	Using rescaled marginal and conditional probability densities of 
$\varrho^{(\powN)}=|\psi^{(\powN)}|^2$ we rewrite the rescaled energy functional 
\Ref{qCOULOMBenergyRESCALED} as a symmetric $N-2$-point average of a two-body energy functional which is 
evaluated with a $S_N$-symmetric two-point probability density \emph{conditioned on the remaining $N-2$ points}.

        More explicitly, invoking the notation stipulated in section \ref{sec:BornEns}, we 
write $(q_1,q_2)=:q^{(2)}\in\Rset^6$ and $(q_3,...,q_N)=:q^{(\powN-2)}\in\Rset^{3(N-2)}$.
	By $\varrho^{(\powN)}_{\powN-2}$ we denote the $N-2$-th marginal probability density of 
$\varrho^{(\powN)}$, i.e.
$\varrho^{(\powN)}_{N-2}(q^{(\powN-2)}):= \int \varrho^{(\powN)}(q^{(\powN)})\dd^{^{6}}\!q$.
	By $\varrho_{[2]}^{(\powN)}$ we denote the conditional two-point probability density obtained 
from $\varrho^{(\powN)}$, i.e. 
$
\varrho_{[2]}^{(\powN)}(q^{(2)}|q^{(\powN-2)})
:= \varrho^{(\powN)}(q^{(\powN)})/\varrho^{(\powN)}_{N-2}(q^{(\powN-2)})
$.
        Now define the two-body energy functional
\begin{equation}
  \pzcE_{\BO,\lambda,\kappa}^{(2)}\bigl(\varrho^{(2)}\bigr) 
=\label{rescaledEtwoN}
	{\textstyle{\frac{\hbar^2}{8m}}} \pzcF^{(2)}\bigl(\varrho^{(2)}\bigr) 
+ z^2e^2\!\!
	\int\!\!\Big(\!-\lambda \big({\textstyle{\frac{1}{|q^{}_1|} +\frac{1}{|q^{}_2|}}}\big)
     + \kappa  {\textstyle{\frac{1}{|q^{}_1-q^{}_2|}}}\!\Big)\varrho^{(2)}(q^{(2)})\dd^{^{6}}\!\!q.
\end{equation}
	This functional is of the Born--Oppenheimer Helium-type; when $\kappa=1/2$ and $\lambda=1$ it becomes 
\emph{the}  Born--Oppenheimer Helium energy functional rescaled to  ``normal form;'' see \cite{Thirring}.
	Note that by scaling $q\to \lambda^{-1}q$ one can always accomplish ``normal form,'' viz.
$\pzcE_{\BO,\lambda,\kappa}^{(2)}\bigl(\varrho^{(2)}\bigr) 
=
\lambda^2 \pzcE_{\BO,1,\kappa/\lambda}^{(2)}\bigl(\varrho^{(2)}\bigr) 
$.

	We are now ready to express the rescaled energy functional \Ref{qCOULOMBenergyRESCALED} 
in terms of \Ref{rescaledEtwoN} and $\varrho^{(\powN)}$.
	For $N=2$, we obviously have 
$
\pzcE_{\BO,\lambda}^{(2)}\bigl(\varrho^{(2)}\bigr) 
=
 \pzcE_{\BO,\lambda,1/2}^{(2)}\bigl(\varrho^{(2)}\bigr)
$, while for $N>2$ we have\footnote{The
		redundant square brackets inside the integral are meant to facilitate the parsing of the formula.}
\begin{equation}
	{\textstyle\frac{1}{N}} \pzcE_{\BO,\lambda}^{(\powN)}\bigl(\varrho^{(\powN)}\bigr) 
=\label{rescaledErewriteASave}
	\int_{\Rset^{3(N-2)}} 
		\Big[\tfrHALF\pzcE_{\BO,\lambda,\frac{\powN-1}{\powN}}^{(2)}
			\bigl(\varrho_{[2]}^{(\powN)}(\,\cdot\,|q^{(\powN-2)})\bigr)\! 
		\Big]\varrho^{(\powN)}_{N-2}(q^{(\powN-2)})\dd^{^{3(N-2)}}\!\!\!\!q.
\end{equation}	
	With \Ref{rescaledErewriteASave} the ground state problem of the $N$-body Hamiltonian \Ref{HamiltonianATOM} 
reduces to the discussion of averages of Born--Oppenheimer Helium-type operators.

        Having \Ref{rescaledErewriteASave}, we now use that $N\mapsto 1-1/N$ is increasing, and that the
set of two-point conditional probability densities of $S^{}_N$-symmetric $N$-point densities is decreasing 
in $N$ in the sense of set-theoretic inclusion.
\qed
\begin{rem}
	Proposition \ref{prop:EmonoUPion} and its proof are essentially contained in a recent paper by 
Hogreve \cite{Hog}, who proved for neutral Born--Oppenheimer atoms that $\cE_{\BO}^{\Bose}(N,N)/N^3$ 
grows monotonically in $N\geq 1$.
	Inspection of Hogreve's proof, see his formulas (7)--(10), reveals that it generalizes verbatim to 
Born--Oppenheimer ions, without restrictions on $\lambda$, and that our proof above (when $\lambda=1$) 
is just Hogreve's argument, recast in the probability measure setting.
\end{rem}
        Note that the restriction $N\geq N^*$ was not needed in our proof of Proposition \ref{prop:EmonoUPion}; 
it enters next.

        We now turn to the last part of Theorem \ref{thm:HARTREElimitSTATES} and 
show that the limit \Ref{vepsLIMlambda} exists, and that it is negative, strictly decreasing, and concave for 
$\lambda>\lambda_*$.
	But armed with Prop. \ref{prop:EmonoUPion} this is easy.
        Namely, as is well-known, the ground state energy $\cE_{\BO}^{\Bose}(\lambda N,N)$ is a negative, 
strictly decreasing, and concave function of  $\lambda>\lambda_*$ when $N\geq N^*$ (cf. \cite{Thirring}).
        Furthermore, by our discussion of the upper Hartree bound on $N^{-3}\cE_{\BO}^{\Bose}(\lambda N,N)$ we have, 
uniformly in $N$, 
\begin{equation}
	N^{-3}\cE^{\Bose}_{\BO}(\lambda N,N)
\leq \label{upperHARTREEestimRHO}
	\inf\big\{ \frake_{\BO,\lambda}(\rho)| \rho\in\Psp(\Rset^3)\cap\{ \surd{\rho}\in\Hsp^1(\Rset^3)\}\big\},
\end{equation}
and it is easy to see that the infimum of the right-hand side is negative and strictly decreasing
$\downarrow -\infty$ as $\lambda\uparrow\infty$.
	Pairing this with Proposition \ref{prop:EmonoUPion} gives us
\begin{cor}
\label{coro:limitEdurchNhochDREI}
	The limit $\veps_\infty(\lambda)=\lim_{N\uparrow\infty}N^{-3}\cE_{\BO}^{\Bose}(\lambda N,N)$ exists.
	Furthermore, $\veps_\infty(\lambda)$ is negative, strictly decreasing, and concave.
\end{cor}

	Of course, by Corollary \ref{coro:limitEdurchNhochDREI} and \Ref{upperHARTREEestimRHO}, also
$\veps_\infty(\lambda)$ inherits the upper bound
\begin{equation}
	\veps_\infty(\lambda)
\leq \label{upperHARTREEestimVEPS}
	\inf\big\{ \frake_{\BO,\lambda}(\rho)| \rho\in\Psp(\Rset^3)\cap\{ \surd{\rho}\in\Hsp^1(\Rset^3)\}\big\}.
\end{equation}

	We next show that the sequence of marginals
$\{\varrho^{(\powN)}_{\lambda,n}\}^{}_{N \geq \max\{n,N^*\}}$ has weak limit points in
$\Msp^s(\Rset^{3n})\cap\{\surd{\rho_n}\in\Hsp^1(\Rset^{3n})\}$ whenever $\lambda\geq\lambda_*$.	
	To this end we need another auxiliary result which likewise is of interest in its own right.
	Its key element is a super-additivity result by Carlen \cite{Carlen}. 
\begin{prp} 
\label{prp:INFOproperties}
	For any $\varrho^{(\powN)}\in\Psp^s\cap \{\surd{\varrho^{(\powN)}}\in\Hsp^1\}$,  
the map $n\mapsto\pzcF^{(n)}\big(\varrhon^{(\powN)}\big)$ has the following properties:

(A) {\emph{Positivity}}: For all $n\leq N$,
\begin{equation}
	\pzcF^{(n)}\big(\varrhon^{(\powN)}\big)\geq 0;
\end{equation}

(B) {\emph{Monotonic increase}}: If $k<n\leq N$, then
\begin{equation}
\pzcF^{(n)}\big(\varrhon^{(\powN)}\big)
\geq 
\pzcF^{(k)}\big(\varrhok^{(\powN)}\big);
\end{equation} 

(C) {\emph{Super-additivity}}: For $ n = k + \ell \leq N$,
\begin{equation}
\hskip-.2truecm
\pzcF^{(n)}\big(\varrhon^{(\powN)}\big)
 \geq 
\pzcF^{(k)}\big(\varrhok^{(\powN)}\big)
+
\pzcF^{(\ell)}\big(\varrho_\ell^{(\powN)}\big).
\end{equation} 
\end{prp}

\medskip\noindent
{\textit{Proof of Proposition \ref{prp:INFOproperties}:}} 

\noindent
Positivity is obvious. Super-additivity was proved in
\cite{Carlen} (his Thm. 3). Monotonic increase now follows from super-additivity and positivity. \qed


	With the help of Proposition \ref{prp:INFOproperties} we are now able to establish
\begin{lem}
\label{lem:COMPACTinHone}
	For any $\lambda\geq\lambda_*$ and $n\in \Nset$ the sequence 
$\{\surd{\varrho^{(\powN)}_{\lambda,n}}\}^{}_{N\geq \max\{n,N^*\}}$ is weakly compact in $\Hsp^1(\Rset^{3n})$.
\end{lem}

\medskip\noindent
{\textit{Proof of Lemma \ref{lem:COMPACTinHone}:}} 

\noindent
	Let $N\geq \max\{n,N^*\}$, then $N= \big\lfloor{N/n}\big\rfloor+k$, 
where $\lfloor{N/n}\rfloor$ is the integer part of $N/n$, so $k<n$.
	We now use super-additivity 
(property $(C)$ in Prop.\ref{prp:INFOproperties}) and then positivity 
(property $(A)$ in Prop.\ref{prp:INFOproperties}) of Fisher information to obtain
\begin{eqnarray} 
	\pzcF^{(\powN)}\big({\varrho}^{(\powN)}_\lambda\big) 
\!\!&\geq&\!\!
	\left\lfloor{\tst{\frac{N}{n}}}\right\rfloor
	\pzcF^{(n)}\big(\varrho^{(\powN)}_{\lambda,n}\big) 
	+
	\pzcF^{(k)}\big(\varrho^{(\powN)}_{\lambda,k}\big) 
\nonumber \\
\!\!&\geq&\!\! \label{superADDestimateBORNagain}
	\left\lfloor{\tst{\frac{N}{n}}}\right\rfloor
	\pzcF^{(n)}\big(\varrho^{(\powN)}_{\lambda,n}\big) ,
\end{eqnarray} 
cf. the Gibbs entropy estimates in \cite{MesserSpohn}, \cite{KieCPAM}.
	(Alternately, \Ref{superADDestimateBORNagain} follows by first using increase (property $(B)$ in 
Prop.\ref{prp:INFOproperties}) to estimate $\pzcF^{(\powN)}\big({\varrho}^{(\powN)}_\lambda\big)$ in terms of
the Fisher information of the 
$n\big\lfloor{N/n}\big\rfloor$-marginal of ${\varrho}^{(\powN)}_\lambda$, then 
super-additivity (C) to get to the $n$-th marginal.)
	Dividing \Ref{superADDestimateBORNagain} by $N$ gives
\begin{equation}
	{\tst{\frac{1}{N}}}\left\lfloor{\tst{\frac{N}{n}}}\right\rfloor
	\pzcF^{(n)}\big(\varrho^{(\powN)}_{\lambda,n}\big) 
\leq\label{preHoneBOUNDn}
	{\tst{\frac{1}{N}}}
	\pzcF^{(\powN)} \big({\varrho}^{(\powN)}_{\lambda}\big)  \quad \forall n\in\Nset.
\end{equation}
	
	We now invoke the virial theorem, which for \emph{any} $\lambda > \Lambda_*(N)$, $N\geq N^*$, gives us
\begin{equation}
	z^2e^2 \pzcV_\lambda^{(\powN)}\!\big(\varrho^{(\powN)}_\lambda\big) 
= \label{virialCOUL}
	-{\textstyle{\frac{\hbar^2}{4m}}} \pzcF^{(\powN)}\!\big(\varrho^{(\powN)}_\lambda\big) ,
\end{equation}
and therefore
\begin{equation}
	\pzcF^{(\powN)}\!\big(\varrho^{(\powN)}_\lambda\big) 
= \label{FISHERisminusEnull}
	-{\textstyle{\frac{8m}{\hbar^2}}}
	\pzcE_{\BO,\lambda}^{(\powN)}\bigl(\varrho^{(\powN)}_\lambda\bigr) .
\end{equation}
	By Proposition \ref{prop:EmonoUPion} (also recalling our rescaling which leads to 
\Ref{qCOULOMBenergyRESCALED}), the map 
$N\mapsto N^{-1}\pzcE_{\BO,\lambda}^{(\powN)}\bigl(\varrho^{(\powN)}_\lambda\bigr)$
is \emph{monotonic increasing}, and so, by \Ref{FISHERisminusEnull}, we conclude that
$N\mapsto N^{-1}\pzcF^{(\powN)}\!\big(\varrho^{(\powN)}_\lambda\big)$ is \emph{monotonic decreasing}.
	Thus,\footnote{Clearly, $\varrho^{(1)}_\lambda$ is the minimizer of a Born--Oppenheimer hydrogenic ion.
        Thus, the \emph{kinetic energy per particle} of the ionic ground state not only decreases 
	with increasing numbers of electrons, it is in fact rigorously bounded by the explicitly computable 
	kinetic energy of an ``electron'' in the hydrogenic ground state.}
\begin{equation}
	N^{-1}\pzcF^{(\powN)}\!\big(\varrho^{(\powN)}_\lambda\big) 
\leq \label{EnullISminusFISHER}
	\pzcF\!\big(\varrho^{(1)}_\lambda\big),
\end{equation}
and so, by \Ref{preHoneBOUNDn},
\begin{equation}
	{\tst{\frac{1}{N}}}\left\lfloor{\tst{\frac{N}{n}}}\right\rfloor
	\pzcF^{(n)}\big(\varrho^{(\powN)}_{\lambda,n}\big) 
\leq\label{almostHoneBOUNDn}
	\pzcF\!\big(\varrho^{(1)}_\lambda\big),
\end{equation}
for $N\geq \max\{n,N^*\}$.
	But since $\frac{1}{N}\big\lfloor\frac{N}{n}\big\rfloor\to{\frac{1}{n}}$ as $N\to\infty$, 
we in fact have
\begin{equation}
	\pzcF^{(n)}\big(\varrho^{(\powN)}_{\lambda,n}\big) 
\leq\label{HoneBOUNDn}
	C(n)\pzcF\!\big(\varrho^{(1)}_\lambda\big),
\end{equation}
for $N\geq \max\{n,N^*\}$.
	With \Ref{kinErhoNfctl}, this means that $\{\surd{\varrho^{(\powN)}_{\lambda,n}}\}^{}_{N\geq \max\{n,N^*\}}$ 
is weakly compact in the homogeneous Sobolev space $\dot\Hsp^1(\Rset^{3n})$.
	On the other hand, we also have that $\{\varrho^{(\powN)}_{\lambda,n}\}^{}_{N\geq \max\{n,N^*\}}$ 
is in $\Psp^s(\Rset^{3n})$, and so the sequence $\{\surd{\varrho^{(\powN)}_{\lambda,n}}\}^{}_{N\geq \max\{n,N^*\}}$ 
is weakly compact in $\Hsp^1(\Rset^{3n})$.
\qed
\smallskip

	By the familiar Sobolev embedding, Lemma \ref{lem:COMPACTinHone} implies:
\begin{cor}
\label{cor:COMPACTinLp}
	For any $\lambda\geq\lambda_*$ and $n\in \Nset$ the sequence 
$\{\varrho^{(\powN)}_{\lambda,n}\}^{}_{N\geq \max\{n,N^*\}}$ is weakly compact in 
$\Lsp^\wp(\Rset^{3n})$ for all $1\leq \wp\leq 3n/(3n-2)$.
\end{cor}

	Clearly, by Corollary \ref{cor:COMPACTinLp} the sequence 
$N\mapsto \varrhon^{(\powN)}$ for $N\geq  \max\{n,N^*\}$
is weakly compact in the set of permutation-symmetric measures $\Msp^s(\Rset^{3n})$ with mass $\leq 1$,
and which are absolutely continuous w.r.t. Lebesgue measure.
	We will now show that any weak limit point $\dotmu_{\lambda,n}$
is in fact also in $\Psp^s(\Rset^{3n})$, i.e. has mass $=1$.
	This will follow from a characterization of the weak limit points in terms of an $N=\infty$ counterpart 
of the finite-$N$ minimum quantum energy principle \Ref{vpErescaled}.
	It will vindicate all the remaining claims in Theorem \ref{thm:HARTREElimitSTATES}.

	We introduce $\Psp^s((\Rset^3)^\Nset)$, the permutation-symmetric probability measures 
on the set of infinite exchangeable sequences in $\Rset^3$.
	Let $\{\mun\}^{}_{n\in\Nset}$ denote the sequence of marginals of any 
$\mu\in{\Psp}^s((\Rset^3)^\Nset)$. 
	The de Finetti \cite{deFinetti} - Dynkin \cite{Dynkin} - Hewitt--Savage \cite{HewittSavage}
decomposition theorem for $\Psp^s((\Rset^3)^\Nset)$ 
states that for each $\mu \in \Psp^s((\Rset^3)^\Nset)$ there exists a unique probability measure 
$\varsigma(\dd\rho|\mu)$ on $\Psp(\Rset^3)$, such that for each $n\in \Nset$,
\begin{equation}
	\mun 
=\label{deFinettiDECOMPconfig}
	\int_{\Psp(\Rset^3)}\rho^{\otimes n} \,\varsigma(\dd\rho|\mu),
\end{equation}
where $\rho^{\otimes n}(\ddn{q}) \equiv \rho(\dd{q})\otimes \cdots\otimes\rho(\dd{q}), n$-fold.
	Note that \Ref{deFinettiDECOMPconfig} expresses the extreme point decomposition of the convex set 
$\Psp^s((\Rset^3)^\Nset)$, see \cite{HewittSavage}.
	If $\mu$ is absolutely continuous w.r.t. Lebesgue measure, one can identify 
the measures $\mu$, $\mu_n$, and $\rho$ with their densities; this notational convenience
should not cause any confusion.
	We also introduce the subsets $\Psp^s_{\wp}((\Rset^3)^\Nset)\subset\Psp^s((\Rset^3)^\Nset)$,
with $\wp\geq 1$, as the probability measures on $(\Rset^3)^\Nset$ whose decomposition measure is 
concentrated on $\Psp_\wp(\Rset^3):=(\Psp\cap\Lsp^\wp)(\Rset^3)$, which are the absolutely 
continuous (w.r.t. $\dd^3{q}$) probability measures on $\Rset^3$ whose density is also in $\Lsp^\wp(\Rset^3)$
(of course, the wording is redundant if $\wp=1$).
	Lastly, we introduce the abbreviation $\mu(g):=\int\!\! g \dd\mu$ for the $\mu$-expectation value of $g$.

	For any $\mu\in \Psp^s_3((\Rset^3)^\Nset)$ we now define the \emph{mean Volta energy of} $\mu$ as
\begin{equation}
\uli\pzcV_\lambda(\mu)
:= \label{meanCONFIGenergyDEF}
\lim_{n\to\infty} 
\ {\tst{\frac{1}{n}}} \mun\!
	\Big(\!-\lambda\!\!{\textstyle\sum\limits_{\sss 1\leq k\leq n}^{} {\frac{1}{|q_k|}}}
     + \tfrn \;{\textstyle\sum\sum\limits_{\hskip-.6truecm \sss 1 \leq k < l \leq n}^{} \frac{1}{|q_k-q_l|}} 
	\Big).
\end{equation}
	Clearly, by the permutation symmetry of $\mu$, this definition yields right away
\begin{equation}
\uli\pzcV_\lambda(\mu)
=\label{meanCONFIGenergy}
-\lambda \muEINS\big(|q|^{-1} \big) + \tfrhalf\muZWEI\big(|q-q'|^{-1} \big),
\end{equation}
and by the linearity of $\mu\mapsto  \uli\pzcV_\lambda\big(\mu\big)$, the presentation 
\Ref{deFinettiDECOMPconfig} now yields
\begin{equation}
	\uli\pzcV_\lambda(\mu)
=\label{meanEconfigVIArep}
	\int_{\Psp_3(\Rset^3)} \Big(-\lambda \rho\big(\!{\textstyle{\,|q|^{-1}}}\big)	+
	\tfrhalf \rho^{\otimes 2}\big({\textstyle{|q-q'|^{-1}}}\big)\Big)\,\varsigma(\dd\rho|\mu).
\end{equation}

	For any $\mu\in \Psp^s((\Rset^3)^\Nset)$ for which each $\surd{\mu_n}\in \Hsp^1(\Rset^{3n})$
we also define the \emph{mean Fisher information of} $\mu$ as limit
\begin{equation}
	\uli\pzcF(\mu)
:=\label{meanFISHERfctl}
	\lim_{n\to\infty} \ {\tst{\frac{1}{n}}} \pzcF^{(n)}\big(\mun\big).
\end{equation}
	Here, $\pzcF^{(n)}\big(\mun\big)$, $n \in\{1,...\}$, is the 
Fisher information of $\mun$, as defined in \Ref{FISHERfctlN}.
	The limit \Ref{meanFISHERfctl} exists or is $+\infty$.
	This is a consequence of the super-additivity and positivity of the Fisher information; see
Proposition \ref{prp:INFOproperties}, 
which holds also with $\varrho^{(\powN)}\in \Psp^s(\Rset^{3\PowN})$ replaced by $\mu\in \Psp^s((\Rset^3)^\Nset)$.

	An analog of \Ref{meanEconfigVIArep} for the mean Fisher information holds as well, thus
\begin{equation}
	\uli\pzcF(\mu)
= \label{FaffineREP}
	\int \pzcF(\rho)\,\varsigma(\dd\rho|\mu).
\end{equation}
	This is a consequence of the next proposition, which seems to be a new result.
\begin{prp} 
\label{prp:meanINFORMATIONisAFFINE}
	The mean Fisher functional \Ref{meanFISHERfctl} is affine linear.
\end{prp} 

\medskip\noindent
{\textit{Proof of Proposition \ref{prp:meanINFORMATIONisAFFINE}:}} 

\noindent
	We use that the Fisher functional and the Gibbs entropy functional are related by the heat flow equation. 
	More precisely, let $\mu\in\Psp^s((\Rset^3)^\Nset)$, and suppose that
$\mu_{k+n}(t)$ solves $\partial_t \mu_{k+n}(t) = \Delta^{(k+n)} \mu_{k+n}(t)$ for $t>0$, with 
$\lim_{t\downarrow 0}\mu_{k+n}(t) = \mu_{k+n}$, where $\Delta_{(k+n)}$ denotes the 
Laplacian in $\Rset^{3(k+n)}$.
	Then integration over the $k$ variables in $\Delta_{(k+n)}$, using that
$\Delta_{(k+n)}= \Delta_{(n)}+\Delta_{(k)}$ and noting that $\mu_n = \int \mu_{n+k}\dd^{3k}q$,
shows that also $\partial_t \mu_n(t) = \Delta_{(n)} \mu_n(t)$ with
$\lim_{t\downarrow 0}\mu_n(t) = \mu_n$.
	Since this holds for all $n\in\Nset$ and  $k\in\Nset$, we have a compatible sequence of equations
defining  $\mu(t)\in \Psp^s((\Rset^3)^\Nset)$, solving $\partial_t \mu(t) = \Delta \mu(t)$ for $t>0$, with 
$\lim_{t\downarrow 0}\mu(t) = \mu$. 
	Moreover, by the linearity of the heat flow equation, if $\mu^{(0)}(t)$ and $\mu^{(1)}(t)$ are
two heat flow evolutions in $\Psp^s((\Rset^3)^\Nset)$, then so is 
$\mu^{(\alpha)}(t) = \alpha \mu^{(1)}(t) +(1-\alpha)\mu^{(0)}(t)$ for all $\alpha\in [0,1]$.

	Now we use that for those $\mu\in\Psp^s((\Rset^3)^\Nset)$ for which $\uli\pzcF\big(\mu\big)$
exists, we have
\begin{equation}
\hskip-.2truecm
	{\textstyle\frac{\dd}{\dd t}}\pzcS^{(n)}\big(\mu_n(t)\big)
=\label{dotENTROPYisFISHER}
	\pzcF^{(n)}\big(\mu_n(t)\big)\ \forall\ t>0.
\end{equation} 
	In particular, if $\uli\pzcF\big(\mu^{(\alpha)}\big)$ exists for $\alpha=0$ and $1$, then
$\uli\pzcF\big(\mu^{(\alpha)}\big)$ exists for all $\alpha\in[0,1]$, and then \Ref{dotENTROPYisFISHER} 
holds with $\mu(0)=\mu^{(\alpha)} \in\Psp^s((\Rset^3)^\Nset)$.

	Next, after division by $n$ and taking $n\to\infty$, \Ref{dotENTROPYisFISHER} for
$\mu(0)=\mu^{(\alpha)}$ yields
\begin{equation}
\hskip-.2truecm
{\textstyle\frac{\dd}{\dd t}}\uli\pzcS\big(\mu^{(\alpha)}(t)\big)
=\label{dotENTROPYisFISHERmean}
\uli\pzcF\big(\mu^{(\alpha)}(t)\big)\ \forall\ t>0.
\end{equation} 
	Here, $\uli\pzcS\big(\mu\big) = \lim_{n\uparrow\infty}n^{-1}\pzcS^{(n)}\big(\mu_n\big)$ 
is the mean Gibbs entropy of $\mu$, cf. \cite{MesserSpohn} and \cite{KieCPAM}; it is the
adaptation to the thermodynamic mean-field limit setting of the ``mean entropy of states''
originally introduced in \cite{RobinsonRuelle} for the conventional thermodynamic limit setting
of the classical grand-canonical ensemble; see also \cite{RuelleBOOK}.
	The mean entropy of $\mu$ is well-defined whenever $\inf_{n\in\Nset} n^{-1}\pzcS^{(n)}\big(\mu_n\big)>C$,
in the sense that it might be $+\infty$ (which is the case iff $C=\infty$); also $C=-\infty$ can occur 
(e.g., if $\mu = \delta_{q^{}_*}^{\otimes\Nset}$), but will not occur with our Born ensemble measures.
	Crucially important, straightforward adaptation of the proof of Proposition 3 in \cite{RobinsonRuelle} 
(cf. \cite{KieCPAM}) yields:
\begin{lem} 
\label{lem:meanENTROPYisAFFINE}
	The mean Gibbs entropy functional  $\uli\pzcS\big(\mu\big)$ is affine linear.
\end{lem} 
	Thus, $\uli\pzcS\big(\mu^{(\alpha)}\big) = 
\alpha\uli\pzcS\big(\mu^{(1)}\big) + (1-\alpha)\uli\pzcS\big(\mu^{(0)}\big)$, and so
\begin{equation}
\hskip-.2truecm
	{\textstyle\frac{\dd}{\dd t}}\uli\pzcS\big(\mu^{(\alpha)}(t)\big)
=\label{DOTmeanENTROPYisAFFINE}
	\alpha	{\textstyle\frac{\dd}{\dd t}}\uli\pzcS\big(\mu^{(1)}(t)\big) 
	+ (1-\alpha)	{\textstyle\frac{\dd}{\dd t}}\uli\pzcS\big(\mu^{(0)}(t)\big).
\end{equation} 
	Now using \Ref{dotENTROPYisFISHERmean} with $\alpha=0$ and $1$ to rewrite r.h.s.\Ref{DOTmeanENTROPYisAFFINE},
then equating it to r.h.s.\Ref{dotENTROPYisFISHERmean}, we obtain
\begin{equation}
\hskip-.2truecm
	\uli\pzcF\big(\mu^{(\alpha)}(t)\big)
=\label{DOTmeanFISHERisAFFINEalpha}
	\alpha	\uli\pzcF\big(\mu^{(1)}(t)\big) 
	+ (1-\alpha)\uli\pzcF\big(\mu^{(0)}(t)\big).
\end{equation} 
	Taking the limit $t\downarrow 0$ proves that $\uli\pzcF$ is affine linear for all 
$\mu^{(\alpha)}\in\Psp^s((\Rset^3)^\Nset)$ for which $\uli\pzcF\big(\mu^{(0)}\big)$
and $\uli\pzcF\big(\mu^{(1)}\big)$ exist.
	Proposition 1 is proved. \qed

	Although we won't need it for our proof, 
the listing of properties of mean Fisher information \Ref{meanFISHERfctl} would be incomplete without 
the next Proposition, which is proved by adaptation of a proof in \cite{RobinsonRuelle}, their proposition 4.
\begin{prp} 
\label{prp:FISHERinfoISlowerSEMIconti}
	The functional $\mu\mapsto\uli\pzcF\big(\mu\big)$ is weakly lower semi-continuous.
\end{prp} 

	Finally we define the \emph{mean quantum energy of} $\mu\in \Psp^s_\wp((\Rset^3)^\Nset)$ by
\begin{equation}
	\uli{\pzcE_{\BO,\lambda}}(\mu)
:=\label{meanQenergy}
	{\textstyle{\frac{\hbar^2}{8m}}}\uli\pzcF(\mu) + z^2e^2 \uli\pzcV_\lambda(\mu).
\end{equation}
	By \Ref{FaffineREP} and \Ref{meanEconfigVIArep} we have 
\begin{equation}
	\uli{\pzcE_{\BO,\lambda}}(\mu)
=\label{meanQenergyREP}
	\int_{\Psp(\Rset^3)} 	\frake_{\BO,\lambda}(\rho) \,\varsigma(\dd\rho|\mu),
\end{equation}
where $\frake_{\BO,\lambda}(\rho)$ is the quantum energy functional 
introduced in \Ref{standardHARTREErhoFCTL}.
	It is well-defined for those $\rho\in\Psp(\Rset^3)$ for which 
$\surd{\rho}\in \Hsp^1(\Rset^3)$, for all $\lambda>0$.
	This definition extends by scaling to all $\rho\in\Msp(\Rset^3)$ for which 
$\surd{\rho}\in \Hsp^1(\Rset^3)$, for all $\lambda>0$.
	In all other situations it is defined as $\frake_{\BO,\lambda}(\rho)=\infty$.
	This completes our setup, and we are ready to characterize the weak limit points
of the sequences of marginal Born ensemble measures.

	In the following we again invoke the restriction that $N\geq \max\{n,N^*\}$ for each 
$\lambda\geq\lambda_*$ and each $n\in\Nset$.

	By Lemma \ref{lem:COMPACTinHone} and its Corollary \ref{cor:COMPACTinLp}, we can extract a subsequence, 
denoted $\{\varrho^{\sss(\dot{N}[N])}_{\lambda,n}\}^{}_{_{N\geq\max\{n,N^*\}}} \Psp^s(\Rset^{3n})$,
for which $\{\surd{\varrho^{\sss(\dot{N}[N])}_{\lambda,n}}\}^{}_{_{N\geq\max\{n,N^*\}}}$ 
converges weakly in $\Hsp^1(\Rset^{3n})$, so that
$\{\varrho^{\sss(\dot{N}[N])}_{\lambda,n}\}^{}_{_{N\geq\max\{n,N^*\}}}$ itself
converges weakly in $\Lsp^\wp(\Rset^{3n})$, for all $1\leq \wp\leq 3n/(3n-2)$.
	We set
\begin{equation}
	\lim_{N\to\infty} \varrho^{\sss(\dot\powN[N])}_{\lambda,n}
= \label{BORNlimptsMEAS}
	\dotmu^\flat_{\lambda,n}.
\end{equation}
	As remarked earlier, at this point we only know that 
$\surd{\dotmu^\flat_{\lambda,n}}\in \Hsp^1(\Rset^{3n})$ and that
$\dotmu^\flat_{\lambda,n}\in \Msp^s(\Rset^{3n})$, with 
$\dotmu^\flat_{\lambda,n}(\Rset^{3n})= \dotmu^\flat_{\lambda,1}(\Rset^{3})\leq 1$. 
	However, we can define 
$\dotmu_{\lambda,n}:= \dotmu^\flat_{\lambda,n}/\dotmu^\flat_{\lambda,1}(\Rset^{3})$; 
clearly, $\dotmu_{\lambda,n}(\Rset^{3n})= 1$. 

	We now use \Ref{preHoneBOUNDn}, with $\dot{N}[N]$ in place for $N$, and let $N\to\infty$.
	By the lower semi-continuity of the Fisher information~\cite{BenguriaBrezisLieb},
\begin{equation}
	\liminf_{N\to\infty} \pzcF^{(n)}\big(\varrho^{(\dot{N}[N])}_{\lambda,n}\big) 
\geq\label{semiCONTestimateBORN}
	\pzcF^{(n)}(\dotmu^\flat_{\lambda,n}),
\end{equation}
while $\frac{1}{\dot{N}}\left\lfloor\frac{\dot{N}}{n}\right\rfloor\to{\frac{1}{n}}$, so from \Ref{preHoneBOUNDn}
we obtain
\begin{equation}
	\lim_{N\to\infty} {\tst{\frac{1}{\dot{N}}}}\pzcF^{(\dot{N})} \big({\varrho}^{(\dot{N})}_{\lambda}\big) 
\geq\label{FISHERnLIMinf}
	{\tst{\frac{1}{n}}} \pzcF^{(n)}(\dotmu^\flat_{\lambda,n}), \quad \forall n\in\Nset.
\end{equation}
	Here we used that we already showed that 
$N^{-1}\pzcF^{(\powN)} \big(\varrho^{(\powN)}_{\lambda}\big)$ has a limit; if that were not yet known, we could
simply replace the ``$\lim$'' by ``$\liminf$'' at l.h.s.\Ref{FISHERnLIMinf}.

	Next, permutation symmetry yields the identity
\begin{equation}
{\tst{\frac{1}{\dot{N}}}} {\varrho}^{(\dot{N})}_\lambda \Big({\tst\frac{1}{\dot{N}}}
	 {\tst{{\;\sum\!\!\!\sum\limits_{\hskip-12pt \sss{1\leq j<k\leq \dot{N}}}^{}} }} |q^{}_j-q^{}_k|^{-1} \Big)
=\label{CONFIGenergyBORNrepulsive} 
	(1-{\tst\frac{1}{\dot{N}}}){\tst\frac{1}{n-1}} \varrho^{(\dot{N})}_{\lambda,n}
\Big({\tfrn} {\tst{\;\sum\!\!\!\sum\limits_{\hskip-12pt \sss{1\leq j<k\leq n}^{}} }} |q^{}_j-q^{}_k|^{-1} \Big).
\end{equation}
	Weak lower semi-continuity of the Coulomb kernel, and its positivity together with $n-1<n$, plus
$1/\dot{N}\to 0$, altogether now yield, for each $n\in \Nset$,
\begin{equation}
	\liminf_{\dot{N}\to\infty} \ {\tst{\frac{1}{\dot{N}}}} {\varrho}^{(\dot{N})}_\lambda
\Big(
   {\tst\frac{1}{\dot{N}}}{\tst{\;\sum\!\!\!\sum\limits_{\hskip-12pt \sss{1\leq j<k\leq \dot{N}}}^{}}}|q^{}_j-q^{}_k|^{-1}
\Big)
\geq\label{REPULSIVEenergyBORNperNn}
	{\tfrn} \dot\mu^\flat_{\lambda,n}
	\Big({\tfrn} {\tst{\;\sum\!\!\!\sum\limits_{\hskip-12pt \sss{1\leq j<k\leq n}^{}} }} |q^{}_j-q^{}_k|^{-1} \Big).
\end{equation}

	Again by permutation symmetry, we also have the identity
\begin{equation}
	{\tst{\frac{1}{\dot{N}}}} {\varrho}^{(\dot{N})}_\lambda 
	\Big(\!-\lambda\!\!{\textstyle\sum\limits_{\sss 1\leq k\leq \dot{N}}^{} {\frac{1}{|q_k|}}}\Big)
=\label{CONFIGenergyBORNnucleus} 
	\tfrn \varrho^{(\dot{N})}_{\lambda,n}
\Big(\!-\lambda\!\!{\textstyle\sum\limits_{\sss 1\leq k\leq n}^{} {\frac{1}{|q_k|}}}\Big).
\end{equation}
	This time, weak $\Lsp^\wp$ convergence with $\wp=3$ yields
\begin{equation}
	\lim_{\dot{N}\to\infty} \ {\tst{\frac{1}{\dot{N}}}} {\varrho}^{(\dot{N})}_\lambda
\Big(\!-\lambda\!\!{\textstyle\sum\limits_{\sss 1\leq k\leq n}^{} {\frac{1}{|q_k|}}}\Big)
=\label{CONFIGenergyBORNnucleusperNn} 
	\tfrn\dot\mu^\flat_{\lambda,n}
\Big(\!-\lambda\!\!{\textstyle\sum\limits_{\sss 1\leq k\leq n}^{} {\frac{1}{|q_k|}}}\Big).
\end{equation}

	Altogether, \Ref{FISHERnLIMinf}, \Ref{REPULSIVEenergyBORNperNn}, \Ref{CONFIGenergyBORNnucleusperNn} 
yield, for each $n\in\Nset$, that
\begin{equation}
	\lim_{N\to\infty} 
	{\tst{\frac{1}{N}}}\pzcE_{\BO,\lambda}^{(\powN)}\big({\varrho}^{(\powN)}_\lambda\big) 
\geq\label{liminfEperNgeqEpern}
	\tfrn\pzcE_{\BO,\lambda}^{(n)}\big(\dot\mu^\flat_{\lambda,n}\big) 
\end{equation}
for \emph{any} weak limit point $\dot\mu^\flat_{\lambda,n}$.
	Here we used that we already showed that 
$N^{-1}\pzcE^{(\powN)}_{\BO,\lambda}\big(\varrho^{(\powN)}_{\lambda}\big)$ has a limit,
namely $\veps_{\infty}(\lambda)$;  otherwise, we could
simply replace $N$ by $\dot{N}[N]$ and ``$\lim$'' by ``$\liminf$'' at l.h.s.\Ref{liminfEperNgeqEpern}.
	As for r.h.s.\Ref{liminfEperNgeqEpern}, scaling yields
\begin{equation}
	\pzcE_{\BO,\lambda}^{(n)}\big(\dot\mu^\flat_{\lambda,n}\big) 
=\label{EperngeqEpernINFLATED}
	\dotmu^\flat_{\lambda,1}(\Rset^{3})
	\pzcE_{\BO,\lambda}^{(n)}\big(\dot\mu_{\lambda,n}^{}\big).
\end{equation}
	Thus, for \emph{any} subsequence $\dot{N}[N]$ such that \Ref{BORNlimptsMEAS} holds, we have
\begin{equation}
	\veps_\infty(\lambda)
\geq\label{liminfEperNgeqEpernINFLATED}
	\dotmu^\flat_{\lambda,1}(\Rset^{3})
	\tfrn\pzcE_{\BO,\lambda}^{(n)}\big(\dot\mu_{\lambda,n}\big)\ \forall\;n\in\Nset.
\end{equation}

	Next, $\{\dot\mu_{\lambda,n}|n\in\Nset\}$, with each $\surd{\dot\mu_{\lambda,n}}\in\Hsp^1$, 
is a compatible sequence of marginal probability measures which defines a
$\dot\mu_\lambda\in \Psp^s((\Rset^3)^\Nset)$.
	Thus, taking the supremum over $n$ of \Ref{liminfEperNgeqEpernINFLATED}, which is equivalent 
to taking $n\to\infty$, recalling the definition \Ref{meanQenergy} of $\uli\pzcE_{\BO,\lambda}$
and its extremal decomposition \Ref{meanQenergyREP}, we obtain
\begin{equation}
	\veps_\infty(\lambda)
\geq\label{liminfEofNrepBORNdot}
	\dotmu^\flat_{\lambda,1}(\Rset^{3})
	\int \frake_{\BO,\lambda}(\rho) \,\varsigma(\dd\rho|\dot\mu_\lambda),
\end{equation}
for {any} subsequence $\dot{N}[N]$ such that \Ref{BORNlimptsMEAS} holds.
	But
\begin{equation}
	\int \frake_{\BO,\lambda}(\rho) \,\varsigma(\dd\rho|\dot\mu_\lambda)
\geq\label{REPeBIGGERinfe}
	\inf \{ \frake_{\BO,\lambda}(\rho) |\rho\in\Psp; \surd{\rho}\in\Hsp^1\},
\end{equation}
so we finally obtain
\begin{equation}
	\veps_\infty(\lambda)
\geq\label{liminfEofNBORN}
	\dotmu^\flat_{\lambda,1}(\Rset^{3})
	\inf \{ \frake_{\BO,\lambda}(\rho) |\rho\in\Psp; \surd{\rho}\in\Hsp^1\}.
\end{equation}

	We now put everything together.
	First of all, the estimates \Ref{liminfEofNBORN} and \Ref{upperHARTREEestimVEPS} together imply that
\begin{equation}
	\veps_\infty(\lambda)
\geq\label{vepsBIGGERveps}
	\dotmu^\flat_{\lambda,1}(\Rset^{3})
	\veps_\infty(\lambda),
\end{equation}
but since $\veps_\infty(\lambda)<0$ by Corollary \ref{coro:limitEdurchNhochDREI}, \Ref{vepsBIGGERveps}
implies that $\dot\mu^\flat_{\lambda,1}(\Rset^3)=1$, and so
\begin{equation}
	\lim_{N\to\infty} \varrho^{(\dot\powN)}_{\lambda,n}
= \label{BORNlimpts}
	\dotmu_{\lambda,n}  \ \forall\; n\in\Nset.
\end{equation}
	Having $\dot\mu^\flat_{\lambda,1}(\Rset^3)=1$, 
the estimates \Ref{liminfEofNBORN} and \Ref{upperHARTREEestimVEPS} together furthermore imply
\begin{equation}
	\veps_\infty(\lambda)
=\label{limEofNBORN}
	\inf \{ \frake_{\BO,\lambda}(\rho) |\rho\in\Psp; \surd{\rho}\in\Hsp^1\}.
\end{equation}
	But then, by \Ref{liminfEofNrepBORNdot} and \Ref{REPeBIGGERinfe}, we also have
\begin{equation}
	\int \frake_{\BO,\lambda}(\rho) \,\varsigma(\dd\rho|\dot\mu_\lambda)
=\label{BORNmeanEisINF}
	\inf \{ \frake_{\BO,\lambda}(\rho) |\rho\in\Psp; \surd{\rho}\in\Hsp^1\}
\end{equation}
for \emph{any} weak limit point $\dot\mu_\lambda$ of the rescaled sequence of
Born's ground state ensemble measures.
	It follows that the decomposition measure $\varsigma(\dd\rho|\dot\mu_\lambda)$ is
supported on those $\rho\in\Psp$ which actually minimize $\frake_{\BO,\lambda}(\rho)$; 
for suppose not, then the average 
$\int \frake_{\BO,\lambda}(\rho) \,\varsigma(\dd\rho|\dot\mu_\lambda)
>
	\inf \{ \frake_{\BO,\lambda}(\rho) |\rho\in\Psp; \surd{\rho}\in\Hsp^1\}$,
in contradiction to \Ref{BORNmeanEisINF}.
	Therefore the infimum is in fact attained;
furthermore, since $\surd{\dot\mu_n}\in\Hsp^1$, we can conclude that the minimizer is actually 
in $\Psp\cap\{ \surd{\rho}\in\Hsp^1\}$. 

	Lastly, by the well-known convexity of $\rho\mapsto\pzcF(\rho)$ (see \cite{BenguriaBrezisLieb}),
the minimizer is unique. 
	By the weak compactness of the sequences of marginals this in turn implies that all subsequences
converge to the same limit.

	This proves Theorem \ref{thm:HARTREElimitSTATES}.
\qed

	This proves Theorem \ref{thm:HartreeC}.
\qed

\subsection{Proofs of Theorem \ref{thm:HartreeLIMITrho}}
	Theorem \ref{thm:HartreeLIMITrho} 
becomes a corollary to the proof of Theorem \ref{thm:HARTREElimitSTATES}, once we add the random variables
into the story. 
	This step supplies proper meaning to the notion of $|\psi^{(\powN)}|^2$ as density of a 
\emph{joint probability measure of an $N$-body system}, which otherwise would be just another name 
for a positive function which integrates to unity.
	As Born emphasized, it is this step which relates $\psi^{(\powN)}$ to the empirical ``dots on the screen''
gathered in laboratory experiments.
	An even more profound insight into the physics is obtained by thinking 
of $|\psi^{(\powN)}|^2$ as density of a \emph{typicality measure}. 
	In this section we pursue both interpretations of $\psi^{(\powN)}$.
        Also here we have to impose that $N\geq N^*$.

\subsubsection{Probability}

	So we finally enter the probabilistic meaning of Born's ensemble for the rescaled 
ground state $\psi^{(\powN)}_\lambda$ of the Hamiltonian \Ref{HamiltonianATOM}, namely
as a family $\{Q^{(\powN)}_k|k\in\Nset\}$ of i.i.d. copies of a random vector $Q^{(\powN)}\in \Rset^{3\PowN}$  
with normalized, stationary\footnote{The density $|\psi^{(\powN)}_\lambda|^2$ is stationary in the sense that
	the wave function $\psi^{(\powN)}_\lambda\!\in\!\Lsp^2(\Rset^{3\PowN})$~is stationary under the 
	action of the unitary group generated by the rescaled Hamiltonian \Ref{HamiltonianATOM}.}
ensemble probability density $\varrho^{(\powN)}_\lambda = |\psi^{(\powN)}_\lambda|^2$. 
        Physically it is more natural to think of the vector $Q^{(\powN)}\in \Rset^{3\PowN}$ as a 
random set of points  $\{Q^{}_1,...,Q^{}_N\} \subset\Rset^3$; i.e. as a representative of
the equivalence class of such vectors under the permutation group $S_N$,  also denoted by $Q^{(\powN)}$.
	Associated to each random set of $N$ points $Q^{(\powN)}$ is a unique normalized empirical one-point 
random ``density''
\begin{equation}
	\uli\Delta^{(1)}_{Q^{(\powN)}}(s) 
= \label{normalEMPmeasONE}
	{\textstyle{\frac{1}{N}} \sum\limits_{\sss 1\leq j\leq N}^{}} \delta_{Q^{}_j}(s),
\end{equation}
and more generally a normalized empirical random $U$-statistic 
$\uli\Delta^{(n)}_{Q^{(\powN)}}$ defined in \Ref{normalEMPmeasUn}.
	The quantities of interest to physics are of the form
\begin{equation}\label{physPROBnormalEMPmeasDISTfeq}
	\Prob\big(\dKR\big(\uli\Delta^{(n)}_{Q^{(\powN)}},\rho^{(n)}\big) > \delta\big) ,
\end{equation}
where ``$\Prob$'' refers to $|\psi^{(\powN)}_\lambda|^2$ as probability density for $Q^{(\powN)}\in\Rset^{3\PowN}$,
and $\rho^{(n)}(s_1,...,s_n) \in (\Psp^s\cap \Csp^0_b)(\Rset^{3n})$ 
is some continuum approximation to a physically meaningful \emph{macrostate} (see below);
in practical situations, $n=1$ or $2$.
          Furthermore, $\dKR$ is some Kantorovich-Rubinstein metric quantifying the weak topology on $\Psp^s(\Rset^{3n})$.

	We inject that we have the luxury of being allowed to remain
somewhat vague about both, the notion of macrostate in general, and which Kantorovich-Rubinstein metric we want to use.
        Namely, since we are invoking the physicists' notion of ``ideal position measurements,'' it is immaterial how 
approximately to these one chooses a ``macrostate'' in real life, and we can allow \emph{any} (piecewise) continuous
$n$-point function as ``macrostate;'' for more on conventional macrostates, see \cite{GoldsteinLebowitz}.
        And since the $N$-body ground state wave function decays exponentially in $\Rset^{3\PowN}$ \cite{Agmon},
and since the same is true in $\Rset^3$ for the Hartree minimizer \cite{BenguriaBrezisLieb}, when
paired with our Theorem \ref{thm:HartreeC} 
 we can conclude that the exponential decay is uniform in $N$ for any $n$th marginal, 
so that the particular choice amongst the standard cost functions in the KR metric is of no essential importance.

        For any fixed $n<N$, with $N\geq N^*$, Born's ensemble entails the conventional \emph{law of large numbers}.
        Namely, let $\cN$ be the number of ideal measurements of $Q^{(N)}$ that are being performed, with $N\geq N^*$ 
fixed.
        Then the law of large numbers says that when $\cN\to\infty$ the sample mean over the independent and 
indentically distributed (i.i.d.) empirical $N$-point densities converges in probability to the theoretical mean, 
given by 
$|\psi^{(\powN)}_\lambda|^2\big(q^{(\powN)}\big)$; and so, for any $n < N$, with $N\geq N^*$, 
the sample mean over the i.i.d. $n$-point densities converges in probability to their theoretical mean
\begin{equation}
 \varrho^{(\powN)}_{\lambda,n}(q^{}_1,...,q^{}_n)
 = \label{BORNnMEAS}
\int_{\Rset^{3(N-n)}}|\psi^{(\powN)}_\lambda|^2\big(q^{(\powN)}\big)\dd{q}_{n+1}\cdots\dd{q}_{N}.
\end{equation}
        Our Theorem \ref{thm:HartreeC} then adds to this that the theoretical mean
$\varrho^{(\powN)}_{\lambda,n}(q^{}_1,...,q^{}_n)$ takes a simple Hartree product form when $N\to\infty$. 

        Our Theorem \ref{thm:HartreeLIMITrho} 
says something much stronger, namely that, in probability, the outcome of a
\emph{single} ideal measurement of the empirical $n$-point density of an $N$-body system will converge 
to the Hartree product density when $N\to\infty$.
        The number $\cN$ of ideal measurements of the empirical $n$-point density plays no role in the argument.
	Of course, since a single ideal measurement of the empirical $n$-point density of an $N$-body system is
equivalent to $N$ ideal measurements of the individual particle positions, also Theorem \ref{thm:HartreeLIMITrho}
can be phrased as a \emph{law of large numbers}, this time the number $N$ of particles in a single system.

\smallskip\noindent
{\textit{A first proof of Theorem \ref{thm:HartreeLIMITrho}:}}

\noindent
        We recall that,  if $Q^{(\Nset)}$ denotes an infinite random sequence of points in $\Rset^3$ 
which are independently and identically distributed by a single-point probability measure 
$\rho$ (having density $\rho$), then their joint probability density is the infinite product 
$\rho^{\otimes\Nset}$.
        Considering $Q^{(\powN)}$ as a subsequence of $Q^{(\Nset)}$, the conventional \emph{law of large numbers} 
states that the associated sequence of $n$-point empirical measures $\uli\Delta^{(n)}_{Q^{(\powN)}}$
 converges in probability for each $n\in\Nset$; viz. for each $n\in\Nset$:
$\dKR\big(\uli\Delta^{(n)}_{Q^{(\powN)}},\rho^{\otimes n}\big) \withNto 0$, in probability.

        Now, as explained in the proof of Theorem \ref{thm:HARTREElimitSTATES}, 
the $N\to\infty$ limit of Born's ensemble measure yields
$\varrho^{(\powN)}_{\lambda}\withNto \rho^{\otimes\Nset}_\lambda$, an infinite product measure on the
set of infinite sequences $q^{(\Nset)}$. 

        This proves Theorem \ref{thm:HartreeLIMITrho}. 
\qed

        The proof of Theorem \ref{thm:HartreeLIMITrho} 
just given brings the law-of-large-numbers character of Theorem \ref{thm:HartreeLIMITrho} to the fore, but
leaves the interpretation in terms of ``a single measurement of the empirical $n$-point density of an 
$N=\infty$ system'' rather implicit.
        To bring out this aspect of Theorem \ref{thm:HartreeLIMITrho}, we now give a second proof.

\smallskip\noindent
{\textit{A second proof of Theorem \ref{thm:HartreeLIMITrho}:}}

\noindent
        We already emphasized that each $q^{(\powN)}\subset\Rset^3$ is associated with a family of 
normalized empirical $n$-point densities $\{\uli\Delta^{(n)}_{q^{(\powN)}}\}_{n\leq\powN}$.
        Now consider $q^{(\powN)}$ as a finite subsequence of an infinite sequence $q^{(\Nset)}$ of points 
in $\Rset^3$. 
        Each infinite sequence $q^{(\Nset)}\subset\Rset^3$ is associated with a family of sequences
$\{\uli\Delta^{(n)}_{q^{(\powN)}}\}_{\powN\geq\max\{n,N^*\}}\in\Psp^s(\Rset^{3n})$, $n\in\Nset$.
        For each $q^{(\Nset)}$ for which there is a $\rho\in\Psp(\Rset^3)$ such that
$\uli\Delta^{(1)}_{q^{(\powN)}} \withNto \rho$ weakly in the sense of measures,
we can define $\uli\Delta^{(1)}_{q^{(\Nset)}}$ to mean the 
normalized measure $\rho\in\Psp(\Rset^3)$ given by the weak limit of $\uli\Delta^{(1)}_{q^{(\powN)}}$
when $N\to\infty$.
        Each such $\uli\Delta^{(1)}_{q^{(\Nset)}}\equiv\rho$ is just the first member of an 
infinite family of normalized $n$-point measures $\uli\Delta^{(n)}_{q^{(\Nset)}}$, $n\in\Nset$, which are defined
analogously as weak limits of the $n$-point empirical measures $\uli\Delta^{(n)}_{q^{(\powN)}}$, $n\in\Nset$, 
as $N\to\infty$.
        It is easy to see from the definition of $\uli\Delta^{(n)}_{q^{(\powN)}}$ that, inevitably,
$\uli\Delta^{(n)}_{q^{(\Nset)}}= \rho^{\otimes n}$, $n\in\Nset$.

        Now, for each $n\leq N$, with $N\geq N^*$, Born's ensemble measure $|\psi^{(\powN)}_\lambda|^2\ddN{q}$ can 
be identified with a probability measure $\varsigma^{(n|N)}_\lambda$ on the set of normalized $n$-point measures; note that 
$\varsigma^{(n|N)}_\lambda$ lives on the singular subset given by the generic version of \Ref{normalEMPmeasUn}, 
viz. \Ref{normalEMPmeasUn} with ${Q^{(\powN)}}$ replaced by 
$q^{(\powN)}=(q^{}_1,...,q^{}_N)\in\Rset^{3n}$, or rather $\{q^{}_1,...,q^{}_N\}\subset\Rset^3$.
        By the explanations given in the previous paragraph, whenever 
$\varsigma^{(n|N)}_\lambda$ has a limit $\varsigma^{(n)}_\lambda$ as $N\to\infty$, then $\varsigma^{(n)}_\lambda$ 
is a probability measure on the set of normalized $n$-point measures which lives on the subset of these which
are $n$-fold product measures, viz.  $\varsigma^{(n)}_\lambda$ is supported by the set of $\rho^{\otimes n}$. 
        Clearly, all the  $\varsigma^{(n)}_\lambda$, $n\in\Nset$, are uniquely determined by
$\varsigma^{(1)}_\lambda =:\varsigma_\lambda$.
        Note that this is precisely the content the Hewitt--Savage decomposition theorem!
        More to the point, the proof of Theorem \ref{thm:HARTREElimitSTATES} 
shows that the Hewitt--Savage decomposition measure 
$\varsigma(\dd\rho|\mu_\lambda)$ simply \emph{is} the limit $N\to\infty$ of Born's ensemble measure
$|\psi^{(\powN)}_\lambda|^2\ddN{q}$ identified with $\varsigma^{(1|N)}_\lambda$. 

        Thus, the fact, found in the proof of Theorem \ref{thm:HARTREElimitSTATES}, 
that $\varsigma(\dd\rho|\mu_\lambda)$ 
is concentrated on a single $\rho_\lambda$ (which is the unique minimizer of the asymptotic Hartree 
functional), means that the probability, of picking, from Born's ensemble, an empirical configuration 
which differs from $\rho_\lambda$, goes to zero when $N\to\infty$.

This is the content of Theorem \ref{thm:HartreeLIMITrho}. \qed

\subsubsection{Typicality}

     Born's statistical / probabilistic interpretation of $|\psi_\lambda|^2$ inevitably suggests
the interpretation of Theorem \ref{thm:HartreeLIMITrho} in terms of a law  of large numbers. 
     This notion refers to the sample mean over many experimental results becoming sharply distributed.
     Yet, in our second proof of Theorem \ref{thm:HartreeLIMITrho} 
we already saw that we can make sharp statements about the outcome
of a \emph{single measurement} of the $n$-point densities in individual systems with overwhelming likelihood.

     Even more penetrating is the notion of \emph{typicality}, in which 
$|\psi_\lambda^{(\powN)}|^2$ features as having the meaning of typicality measure rather than probability
measure.
     The notion of typicality also refers to individual systems, but it refers to these in their own right, 
without reference to any measurements being performed on them: If the overwhelming amount of configurations
$Q^{(\powN)}$ correspond to the same $\rho^{(n)}_{\mathrm{typ}}$, then this $\rho^{(n)}_{\mathrm{typ}}$ is 
characteristic, i.e. \emph{typical} for the $N$-body system.
	It is this notion which really lies at the heart of Boltzmann's insights \cite{Boltzmann}
about the ergodic ensemble, not the subsequently invented dynamical notion of ``ergodicity;'' 
see \cite{Goldstein}, and also \cite{KieJSTAT}.
	A \emph{typical} $n$ point density 
$\rho^{(n)}_{\mathrm{typ}}(s_1,...,s_n) \in (\Psp^s\cap \Csp^0_b)(\Rset^{3n})$ 
is therefore defined implicitly as the --- in the simplest case: unique --- function for which
\begin{equation}\label{LLNempMEAS}
\Meas\big(\dKR\big({\uli\Delta^{(n)}_{Q^{(\powN)}}},\rho^{(n)}_{\mathrm{typ}}\big) > \delta\big) 
\withNto 0 \quad \forall \delta>0,
\end{equation}
where
``$\Meas$'' refers to $|\psi^{(\powN)}_\lambda|^2$ as typicality measure (density) on $\Rset^{3\PowN}$.

Clearly, everything we said about Theorem \ref{thm:HartreeLIMITrho} 
in terms of probability can be rephrased in terms of typicality.
\qed

\section{Concluding remarks}
\noindent
	In this paper we have presented a novel approach to the Hartree limit of bosonic Born--Oppenheimer
atoms or ions.
	Some of the items listed in our Theorem \ref{thm:HartreeC} have previously been proved with different 
techniques, while other items seem new, indeed.
        Our approach is based on Born's statistical ensemble interpretation of $|\psi|^2$, see Theorems 
\ref{thm:HartreeLIMITrho} and \ref{thm:HARTREElimitSTATES}.
        Their proofs are inspired by an established strategy in classical statistical mechanics, 
pioneered in \cite{MesserSpohn}.
	The key idea is that the role played by the negative Gibbs entropy in the classical petit-canonical ensemble,
is played by the Fisher information functional in Born's quantum ensemble.

	Along the way we also established an auxiliary result, namely the monotonic increase of
$N^{-3}\cE_{\BO}^{\Bose}(\lambda N,N)$ with $N$, see Prop. \ref{prop:EmonoUPion}.
	The monotonicity of $N^{-3}\cE_{\BO}^{\Bose}(N,N)$ (i.e. $\lambda=1$)
had already been shown by Hogreve \cite{Hog}, who used it to establish the existence of a stable 
bosonic ``He$^-$ ion'' in Born--Oppenheimer approximation. 

        Due to our currently incomplete knowledge of the optimal range of $Z$ values for which a proper bosonic ground
state $\Psi^{(\powZ,\powN)}_\Bose$ exists, the proofs of our theorems work under the restriction that $N\geq N^*$, 
where $N^*$ is some finite positive integer.
        In this regard we formulated two reasonable conjectures, 1 and 3, which, if true, would imply that $N^*=1$, 
thereby eliminating the need to restrict $N$ to $N\geq N^*$. 
        Our conjectures are also of some interest in their own right; in particular, as pointed out by the referee,
Conjecture 1 is closely related to a known conjecture, 2, formulated in \cite{LiebSeiringer}.
        It would be very nice if all these conjectures would be settled in some future work.

        As also noted by the referee, the convergence of the states accomplished here is relatively weak, and 
it would be very interesting to know if the technique could be modified to prove the convergence in trace norm of 
the family of reduced density matrices for the ground states. 

	Other projects worthy of our attention in the future include the generalization of our technique to the 
\emph{conditional} ground state ensemble of Galilei-invariant ``bosonic atoms and ions;'' i.e. to drop the 
Born--Oppenheimer approximation and to treat the system in its center-of-mass frame.
        In the same category are ``bosonic neutron stars.''
        Less obvious, and therefore more intriguing, is the generalization to fermionic atoms, ions, and stars. 
In this case one would expect to approach Hartree--Fock theory \cite{LiebSimon}.

	In this work the terminology ``bosonic atom'' invariably meant an atom 
whose ``electrons'' were bosons, and an atom  with true electrons was called ``fermionic atom.''
	Same for ions. 
	While this terminology has been in use by parts of the physics community
for some time by now, the term ``bosonic atom'' can also be found in the physics literature to mean an atom
with conventional electrons, but whose total spin angular momentum quantum number is an integer. 
	The most prominent example, perhaps, is the neutral helium atom ${}^4$He in its ground state, having
spin $0$.
	Experimental studies of $N\gg 1$ such bosonic atoms in a trap have been making headlines in the recent past 
with the empirical observation of Bose--Einstein condensation in gases consisting of bosonic alkali atoms
\cite{BEa,BEb}.

	Since theoretical physicists have employed the Hartree approximation also to study Bose--Einstein
condensation, it is worthwhile to point out that the mathematical approach to the Hartree limit developed in
the previous sections readily extends to the Hamiltonian used by theorists studying BE condensates, which
yields the energy functional \Ref{qCOULOMBenergyRESCALED} with $ze=1$ and with potential energy functional 
\Ref{potErhoNfctl} replaced by
\begin{equation}
	\pzcV_\lambda^{(\powN)}\!\left(\varrho^{(\powN)}\right) 
=\label{potErhoNfctlBE}
	\int\!\Big(\lambda\!\!{\textstyle\sum\limits_{\sss 1\leq k\leq N}^{} |q_k|^2}
     + \tfrN \;{\textstyle\sum\sum\limits_{\hskip-.6truecm \sss 1 \leq k < l \leq N}^{} U(|q_k-q_l|)} 
	\Big)	 \,\varrho^{(\powN)}\dd^{^{3N}}\!\!q;
\end{equation}
here, $\lambda>0$ now represents the strength of the confining harmonic potential, and $U>0$ is some short range,
integrable repulsive potential (frequently even taken to be bounded).
	We leave it to the interested reader to verify that our Hartree limit theorems and proofs extend nearly 
verbatim to this setting; some of the technical estimates even simplify.

\smallskip  
\noindent
{\textbf{Acknowledgment}.} 

\noindent
	Work supported under  NSF grant DMS-0807705.
	Any opinions expressed in this paper are entirely those of the author and not those of the NSF. 
	I thank Shelly Goldstein for his helpful comments on section 3.2, and Brent Young for catching
	some slips of pen in the original manuscript.
	I am indebted to the anonymous, excellent referee whose helpful suggestions improved the presentation,
and whose penetrating comments have helped to eliminate some mistakes in the original version, 
as a result of which Conjectures 1 and 3 were formulated.

\newpage

\newpage

\end{document}